\documentclass[aps,amsmath,preprint, prd,nofootinbib,superscriptaddress,]{revtex4}
\usepackage{graphicx}
\usepackage{color}
\usepackage{amsmath,amssymb}
\usepackage{amsfonts}
\usepackage{mathrsfs}
\usepackage{bm}
\usepackage{mathrsfs}
\usepackage{longtable}
\usepackage{slashed}

\newcommand{\beq}{\begin{eqnarray}}
\newcommand{\eeq}{\end{eqnarray}}

\newcommand{\e}{{\rm e}}
\newcommand{\tr}{\, {\rm tr}\, }

\newcommand{\aaa}{\mathfrak{a}}
\newcommand{\bbb}{\mathfrak{b}}
\newcommand{\eee}{\mathfrak{e}}
\newcommand{\F}{\mathcal{F}}
\newcommand{\A}{\mathcal{A}}
\usepackage{comment}
\usepackage{epstopdf}

\def\simge{\mathrel{%
   \rlap{\raise 0.511ex \hbox{$>$}}{\lower 0.511ex \hbox{$\sim$}}}}
\def\simle{\mathrel{
   \rlap{\raise 0.511ex \hbox{$<$}}{\lower 0.511ex \hbox{$\sim$}}}}
\def\bigs{\mathrel{
   \rlap{\raise 0.531ex \hbox{$>$}}{\lower 0.531ex \hbox{$<$}}}}

\usepackage[normalem]{ulem}  
\newcommand{\com}[1]{{\sf\color[rgb]{0,0,1}{#1}}}

\begin{document}

\begin{flushright}
KEK-TH-1815, RIKEN-QHP-189, RBRC-1134 \\
\end{flushright}

\vspace{0.6cm}

\title{\Large Euler-Heisenberg-Weiss action for QCD+QED}

\author{\bf Sho Ozaki}
\email[E-mail: ]{sho@post.kek.jp}
\affiliation{Theory Center, IPNS, High Energy Accelerator Research Organization (KEK), 
1-1 Oho, Tsukuba, Ibaraki 305-0801, Japan}
\author{\bf Takashi Arai}
\email[E-mail: ]{tarai@post.kek.jp}
\affiliation{Theory Center, IPNS, High Energy Accelerator Research Organization (KEK), 
1-1 Oho, Tsukuba, Ibaraki 305-0801, Japan}

\author{\bf Koichi Hattori}
\email[E-mail: ]{koichi.hattori@riken.jp}
\affiliation{RIKEN BNL Research Center Bldg. 510A, Brookhaven National Laboratory, Upton, New York 11973 USA}
\affiliation{Theoretical Research Division, Nishina Center, RIKEN, Wako, Saitama 351-0198, Japan}

\author{\bf Kazunori Itakura}
\email[E-mail: ]{kazunori.itakura@kek.jp}
\affiliation{Theory Center, IPNS, High Energy Accelerator Research Organization (KEK), 
1-1 Oho, Tsukuba, Ibaraki 305-0801, Japan}
\affiliation{Graduate University for Advanced Studies (SOKENDAI),
1-1 Oho, Tsukuba, Ibaraki 305-0801, Japan}

\begin{abstract}

We derive an analytic expression for one-loop effective action of
QCD+QED at zero and finite temperatures by using the Schwinger proper
time method. The result is a nonlinear effective action 
not only for electromagnetic and chromo-electromagnetic fields but also for the Polyakov loop, and thus reproduces the Euler-Heisenberg action in QED, QCD, and QED+QCD, and also the Weiss potential for the Polyakov loop at finite
temperature. As applications of this ``Euler-Heisenberg-Weiss" action in
QCD+QED, we investigate quark pair productions induced by QCD+QED fields
at zero temperature and the Polyakov loop in the presence of strong
electromagnetic fields. Quark one-loop contribution to the effective
potential of the Polyakov loop explicitly breaks the center symmetry,
and is found to be enhanced by the magnetic field, which is consistent
with the inverse magnetic catalysis observed in lattice QCD simulation.

\end{abstract}


\maketitle


\section{Introduction}

The very first stage in a high-energy heavy-ion collision is dominated by extremely strong {\it chromo-electromagnetic} (chromo-EM) fields reflecting colliding nuclei filled with high-density gluons (color glass condensate). Such a state with strong fields is called a ``glasma'' which is named since it is a transitional state between a color {\it glass} condensate (before the collision) and a quark-gluon {\it plasma} (QGP)~\cite{Lappi:2006fp}. The glasma is characterized by a field strength ${\cal F}$ of the order of the saturation scale: $g{\cal F}\sim Q_{s}^2$ (with $g$ being the QCD coupling). Notice that the saturation scale $Q_s$ is a semihard scale representing a typical transverse momentum of gluons in a colliding nucleus and can become large enough, at high energies, compared to light quark masses $Q_s\gg m_q$. Besides, it has long been known that heavy-ion collisions, with electrically charged nuclei, are accompanied by  {\it electromagnetic} (EM) fields, but only recently was it seriously recognized that the strong EM fields could affect time evolution of heavy-ion collision events since the strength $F$ of the EM fields  could be as large as or even greater than the nonperturbative QCD scale $\Lambda_{\rm QCD}$, namely $eF\simge \Lambda_{\rm QCD}^2$ and thus $eF\gg m_q^2$ \cite{Kharzeev:2007jp, Skokov:2009qp, Bzdak:2011yy, Deng:2012pc}. Since both the chromo-EM and EM fields created in heavy-ion collisions can be strong enough compared with the light quark masses, the effects of strong fields cannot be treated as perturbation (even though the coupling constants are small), but must be treated in a nonperturbative way. Then we expect nonlinear and nonperturbative phenomena associated with the strong fields to occur. Typical examples of such phenomena include particle productions (quarks, antiquarks and gluons) from these strong fields (the Schwinger mechanism), which must be a key towards understanding the formation of QGP.

While the (coherent) chromo-EM fields will disappear as the QGP is formed, the EM fields could survive longer due to Faraday's law, which works in the presence of a conducting medium~\cite{Tuchin:2013ie, Gursoy:2014aka}. If the EM fields survive at a strong enough level until the formation of QGP, and even until the end of the QGP's lifetime, we need to describe the QCD phase transition 
with the effects of strong EM fields taken into account. Notice that the effects of strong {\it magnetic} fields on thermodynamical or fundamental quantities of QGP can be investigated in lattice QCD simulations, and are indeed found to be large. For example, at zero temperature, lattice QCD simulations confirmed the ``magnetic catalysis'' as predicted in several effective models \cite{Gusynin:1994re, Gusynin:1994xp, Kashiwa:2011js, Gatto:2010pt, Kamikado:2013pya, Cohen:2007bt, Andersen:2012dz, Andersen:2012zc} in which the value of chiral condensate increases with increasing magnetic field strength. On the other hand, at finite temperature, lattice QCD simulations almost at the physical point concluded \cite{Bali:2012zg, Bali:2011qj} that the magnetic catalysis does not necessarily occur at all the temperature regions, but rather gets weakened and even shows opposite behavior with increasing temperature. Such behavior of the chiral condensate around the critical temperature is called ``magnetic inhibition'' \cite{Fukushima:2012kc} or ``inverse magnetic catalysis'', which eventually gives rise to decreasing critical temperature.
For recent reviews on the phase diagram of chiral phase transitions in strong magnetic fields, see, e.g., Refs.~\cite{Andersen:2014xxa, Miransky:2015ava}.
Furthermore, it is reported~\cite{Bruckmann:2013oba} that the (pseudo)critical temperature of the confinement-deconfinement phase transition (for the Polyakov loop) also decreases with increasing magnetic field. This is achieved by increasing Polyakov loop expectation values. Probably, these two phenomena are related to each other. However, so far, there is no clear explanation about the physical mechanism behind this (for recent attempts, see Refs.~\cite{Kojo:2012js, Kojo:2013uua} and \cite{Braun:2014fua, Mueller:2015fka}).

We can investigate these two aspects, namely the nonlinear and nonperturbative dynamics of strong fields (including particle production) and the phase transition under strong external fields, within a single framework of an effective action. So far, effective actions for QED and QCD in various external conditions have been extensively explored. First of all, Euler and Heisenberg derived a nonlinear effective action for constant EM fields at the electron's one-loop level, known as the Euler-Heisenberg (EH) action~\cite{Heisenberg:1935qt}. Later, Schwinger reproduced the same action in a field-theoretical manner, which is the so-called Schwinger proper time method~\cite{Schwinger:1951nm}. The EH action at finite temperature is computed in imaginary time formalism~\cite{Dittrich:1979ux, Gies:1998vt} as well as in real time formalism~\cite{Cox:1984vf, Loewe:1991mn}. Furthermore, an analog of the EH action in QCD (for chromo-EM fields) has been evaluated too within a similar method at zero and finite temperatures \cite{Savvidy:1977as, Matinyan:1976mp, Nielsen:1978rm, Leutwyler:1980ma, Schanbacher:1980vq, Dittrich:1983ej, Cea:1987ku, Cho:2002iv, Dittrich:1980nh, Gies:2000dw}. Lastly, the most recent progress was to compute the EH action at zero temperature when both the EM and chromo-EM fields are present, which was done by one of the authors and B.~V.~Galilo and S.~N.~Nedelko independently~\cite{Ozaki:2013sfa, Galilo:2011nh}. The author of Ref.~\cite{Ozaki:2013sfa} used this effective action to investigate the QCD vacuum (gluon condensate) in the presence of strong magnetic fields. Though all of these are about the effective action for strong fields and choromo-EM condensates, it should be possible to include the Polyakov loop at finite temperature. 
Indeed, an effective action (or potential) for the Polyakov loop at the one-loop level was computed independently by D.~J.~Gross, R.~D.~Pisarski, and L.~G.~Yaffe~\cite{Gross:1980br}, and by N.~Weiss~\cite{Weiss:1980rj, Weiss:1981ev}, and the result is called the Weiss potential.
In the present paper, we are going to derive an analog of the EH effective action in QCD+QED at finite temperature with the Polyakov loops included. Thus, the result may be collectively called the ``Euler-Heisenberg-Weiss action." Our result is also a generalization of the one obtained by H.~Gies~\cite{Gies:2000dw}, who computed an effective action for the Polyakov loop and the chromo-electric field.

The paper is organized as follows: In the next section, we will derive the effective action for QCD+QED at finite temperature by using the Schwinger proper time method. 
Variables of the effective action are  the EM and chromo-EM fields as well as the Polyakov loop, and one can reproduce the previous results (the EH action with QCD+QED fields, the Weiss potential, etc.) in various limits. Then, we discuss some applications of our effective action in Sec. III. First, we investigate quark-antiquark pair production in QCD+QED fields at zero temperature. We obtain the quark production rate in the presence of QCD+QED fields, which allows us to study the quark pair production with arbitrary angle between the EM and chromo-EM fields. Next, we study an effective potential for the Polyakov loop with electromagnetic fields. We find that the magnetic field enhances the explicit center symmetry breaking, while the electric field reduces it. This indicates that the (pseudo)critical temperature of the confinement-deconfinement phase transition decreases (increases) with increasing magnetic (electric) field. Finally, we conclude our study in Sec. IV.

\section{one-loop effective action for QCD+QED at finite temperature}


In this section, we derive the one-loop effective action for QCD+QED at finite temperature. 
The effective action will be a function of chromo-EM and EM fields, as well as the Polyakov loop. Notice that both the strong fields and the Polyakov loop can be treated as {\it background fields} so that the background field method is applicable. We will take quantum fluctuations around the background fields up to the second order in the action, and integrate them in the path integral. This corresponds to computing the action at the one-loop level.

We shall begin with the four-dimensional QCD action of the SU$(N_{c})$ gauge group with $N_{f}$ flavor quarks interacting with EM fields:
\beq
S_{\rm QCD+QED}
&=& \int d^{4}x \left\{-\frac{1}{4} F_{\mu \nu}^{a} F^{a \mu \nu} - \frac{1}{4} f_{\mu \nu} f^{\mu \nu} + \bar{q} \left( i \gamma_{\mu} D^{\mu} - M_{q} \right) q\right\} \, ,
\label{QCD+QEDaction}
\eeq
where the covariant derivative contains gluon fields\footnote{Throughout the paper, we use $a,b,c$ (and $h$) for adjoint color indices ($a,b,c=1, \ldots,N_c^2-1$), $i$ for fundamental color indices $(i=1,\ldots,N_c)$, $\mu,\nu,\alpha,\beta$ for Lorentz indices, and $f$ for flavor indices $(f=1,\ldots,N_f)$.} $A^{a}_{\mu}$ $(a=1,\ldots,N_c^2-1)$ and U(1) gauge fields $a_{\mu}$ as
\beq
D_{\mu} = \partial_{\mu} - igA_{\mu}^{a} T^{a} - ieQ_{q}a_{\mu}\, ,
\label{covderivative-all}
\eeq
and the gluon and EM field-strength tensors are given by
$
F_{\mu \nu}^{a}
= \partial_{\mu} A_{\nu}^{a} - \partial_{\nu}A_{\mu}^{a} + gf^{abc} A_{\mu}^{b} A_{\nu}^{c} $ and $
f_{\mu \nu}
= \partial_{\mu} a_{\nu} - \partial_{\nu} a_{\mu}\, ,
$
respectively. In this paper, we treat the EM fields just as background fields, and assume that the field strengths are constant so that $\partial f = 0$. We abbreviate color, flavor, and spinor indices of the quark field in Eq.~(\ref{QCD+QEDaction}).
Mass and charge matrices of quarks are given by $M_{q} = {\rm{diag}}(m_{q_{1}}, m_{q_{2}}, \ldots, m_{q_{N_{f}}} )$ and $Q_{q} = {\rm{diag}}( Q_{q_{1}}, Q_{q_{2}}, \ldots, Q_{q_{N_{f}}} )$. As for the gluon field, we apply the background field method and decompose the gluon field into a slowly varying background field ${\cal A}_{\mu}^{a}$ and a quantum fluctuation $\tilde{A}_{\mu}^{a}$ as
\beq
A_{\mu}^{a} = {\cal A}_{\mu}^{a} + \tilde{A}_{\mu}^{a}\, .
\eeq
Here we employ the covariantly constant field as a background field, which obeys the following condition \cite{Batalin:1976uv, Gyulassy:1986jq, Tanji:2011di}:
\beq
{\cal D}_{\rho}^{ac} {\cal F}^{c}_{\mu \nu} = 0\, ,
\label{CondtionCovariantConstant}
\eeq
where the covariant derivative ${\cal D}_\mu$ is defined only with respect to the gluon background field: 
\beq
{\cal D}_{\mu}^{ac} = \partial_{\mu} \delta^{ac} + g f^{abc} {\cal A}^{b}_{\mu}
\, , \label{covderivative-gluon}
\eeq 
and ${\cal F}_{\mu \nu}^{a} = \partial_{\mu} {\cal A}_{\nu}^{a} - \partial_{\nu} {\cal A}^{a}_{\mu} + g f^{abc} {\cal A}_{\mu}^{b} {\cal A}_{\nu}^{c}$. 
From the condition (\ref{CondtionCovariantConstant}), the field-strength tensor ${\cal F}^{a}_{\mu \nu}$ can be factorized as
${\cal F}^{a}_{\mu \nu}  = {\cal F}_{\mu \nu} n^{a}$, where ${n}^a$ is a unit vector in color space, normalized as ${n}^{a}{n}^{a} = 1$,
whereas ${\cal F}_{\mu \nu}$ expresses the magnitude of the chromo-EM field.
We further assume that ${\cal F}_{\mu \nu}$ is very slowly varying, satisfying $\partial_{\sigma} {\cal F}_{\mu \nu} = 0$, which allows us to obtain the analytic expression of the EH action for QCD, just as in QED. Both ${\cal F}_{\mu \nu}$ and ${n}^a$ are space-time independent. The background field ${\cal A}^a_\mu$ is proportional to the color unit vector ${n}^a$ as
\beq
{\cal A}_{\mu}^{a}
&=& {\cal A}_{\mu} {n}^{a}\, ,
\label{BackgroundField}
\eeq
and the field-strength tensor ${\cal F}_{\mu \nu}$ has an Abelian form, ${\cal F}_{\mu \nu} = \partial_{\mu} {\cal A}_{\nu} - \partial_{\nu} {\cal A}_{\mu}.
$
This background field (\ref{BackgroundField}) indeed satisfies the condition (\ref{CondtionCovariantConstant}).
By using the background field and the quantum fluctuation, the full gluon field-strength tensor can be decomposed as
\beq
F^{a}_{\mu \nu}
&=& {\cal F}_{\mu \nu} {n}^{a} + ( {\cal D}_{\mu}^{ac} \tilde{A}_{\nu}^{c} - {\cal D}_{\nu}^{ac} \tilde{A}_{\mu}^{c}) + gf^{abc} \tilde{A}_{\mu}^{b} \tilde{A}_{\nu}^{c}\, .
\eeq 
Applying the background gauge for the quantum fluctuation,
\beq
{\cal D}^{ac}_{\mu} \tilde{A}^{c}_{\mu} 
&=& 0\, ,
\eeq
we get the gauge fixed action in the presence of EM fields,
\beq
S_{\rm QCD+QED}
&=& \int d^{4}x 
\left[  - \frac{1}{4} \left\{ 
        {\cal F}_{\mu \nu} {n}^{a} 
      + \left( {\cal D}_{\mu}^{ac} \tilde{A}_{\nu}^{c} 
         - {\cal D}_{\nu}^{ac} \tilde{A}_{\mu}^{c} \right) 
      + g f^{abc} \tilde{A}_{\mu}^b \tilde{A}_{\nu}^{c} 
                      \right\}^{2} 
       - \frac{1}{2 \xi} ( {\cal D}^{ac}_{\mu} \tilde{A}^{c \mu} )^{2} \right. 
      \nonumber \\
&& \qquad \quad \left. 
     - \bar{c}^{a} \left( {\cal D}_{\mu} D^{\mu} \right)^{ac} c^{c} 
     + \bar{q}\left(i \gamma_{\mu} D^{\mu} - M_{q} \right) q 
     - \frac{1}{4} f_{\mu \nu} f^{\mu \nu}
\right]\, ,
\eeq
where $c$ is the ghost field and $\xi$ is the gauge parameter. 
Notice that one of the covariant derivatives in the ghost kinetic term $D_{\mu}^{ac}$ and the one in the quark kinetic term $D_{\mu}$ defined in Eq.~(\ref{covderivative-all}) contain all the gauge fields.
The effective action for the background fields ${\cal A}_\mu$ 
and $a_\mu$
can be obtained through the functional integral as 
\beq
{\rm{exp}} \Big( i S_{\rm eff}[{\cal A}_\mu, a_\mu] \Big)
&\equiv& \int {\mathscr D} \tilde{A} {\mathscr D}c {\mathscr D} \bar{c} {\mathscr D}q {\mathscr D} \bar{q} \ \ {\rm{exp}} \left( i \int d^{4}x S_{\rm QCD+QED} \right)\, .
\eeq
We perform the functional integral with fluctuations taken up to the second order. This corresponds to evaluating the one-loop diagrams as shown in Fig.~1. 
\begin{figure}
\begin{minipage}{0.8\hsize}
\begin{center}
\includegraphics[width=1.0 \textwidth]{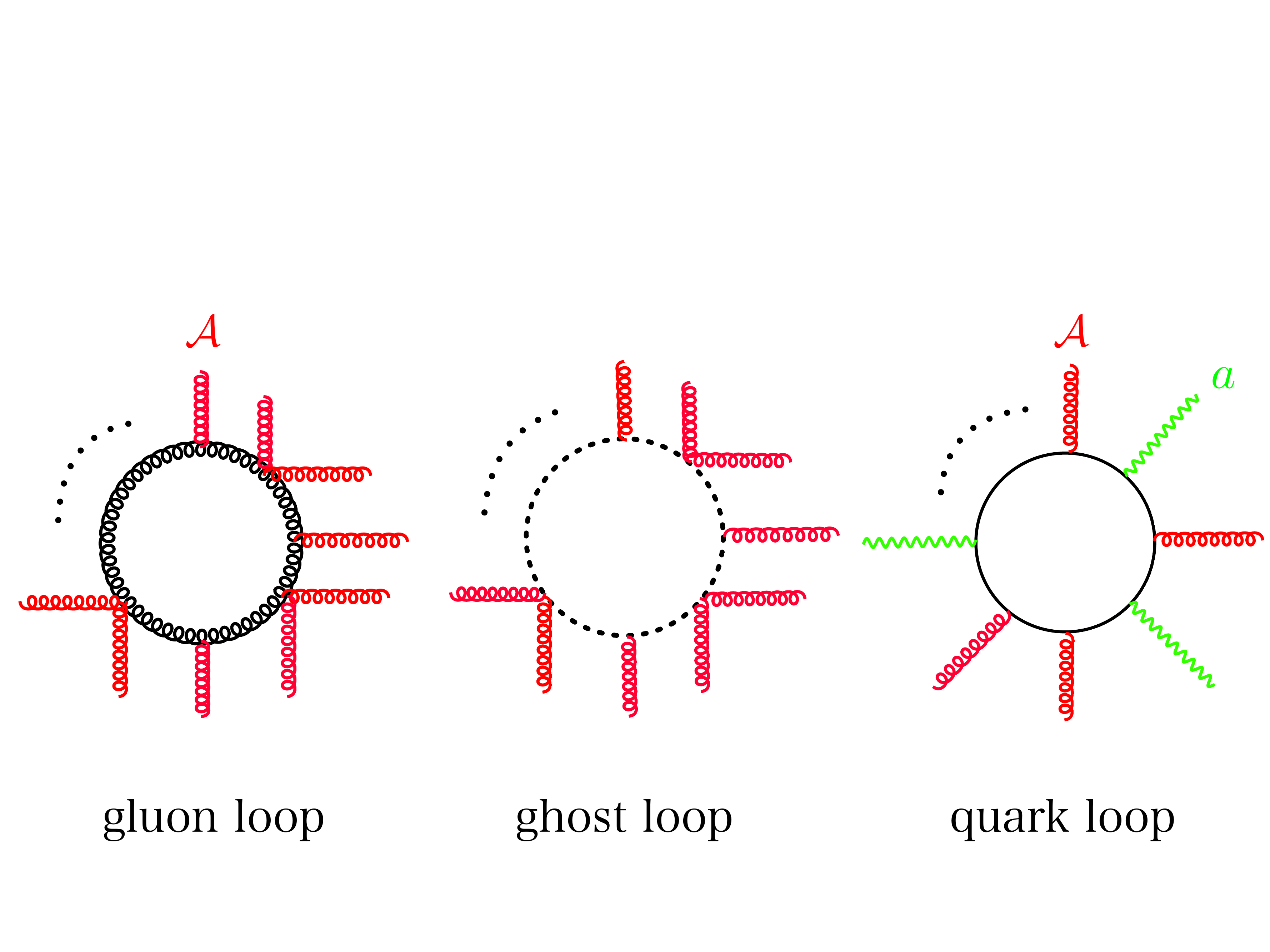}
\vskip -0.1in
\end{center}
\end{minipage}
\caption{
Typical loop diagrams contributing to the effective action. 
The field $\A$ contains both the chromo-EM fields and the Polyakov loop.
}
\end{figure}
The gluon, ghost, and quark loop integrations can be separately done, and one finds, respectively,
\beq
&&\!\!\int\!\! {\mathscr D} \tilde{A}\,  {\rm{exp}}
\left\{ \int \! d^{4}x \frac{-i}{2} \tilde{A}^{a \mu} \left[
- ( {\cal D}^{2})^{ac} g_{\mu \nu} - 2 g f^{abc} {\cal F}^{b}_{\mu \nu} 
\right] \tilde{A}^{c \nu} \right\}
\! ={\rm{det}} \! \left[ - ( {\cal D}^{2})^{ac} g_{\mu \nu} 
                     - 2 g f^{abc} {\cal F}_{\mu \nu}^{b} 
              \right]^{-\frac12}, \nonumber \\
&&\!\!\int\! {\mathscr D} c {\mathscr D} \bar{c} \ {\rm{exp}}
\left\{ i \int d^{4}x \ \bar{c}^{a} \left[ - ( {\cal D}^{2} )^{ac} \right] c^{c} \right\}
= {\rm{det}} \left[ - ( {\cal D}^{2} )^{ac} \right]^{+1}, \label{full_actions} 
\\
&&\!\!\int\! {\mathscr D} q {\mathscr D} \bar{q} \ {\rm{exp}} 
\left\{ i \int d^{4}x \  \bar{q} \left(
i \gamma_{\mu} \hat{\cal D}^{\mu} - M_{q} \right) q \right\}
= {\rm{det}} \left[ i \gamma_{\mu} \hat{\cal D}^{\mu} - M_{q} \right]^{+1}.
\nonumber
\eeq
Here we have taken the Feynman gauge, $\xi = 1$.
In the quark one-loop contribution, the covariant derivative $\hat{\cal D}_{\mu}$ contains both of the background fields ${\cal A}_\mu$ and $a_\mu$:
\beq
\hat{\cal D}_{\mu}
&=& {\cal D}_{\mu} -ieQ_{q} a_{\mu}\nonumber\\
&=& \partial_{\mu} - ig {\cal A}_{\mu}^{a} T^{a} -ieQ_{q} a_{\mu}\, .
\label{CovariantDerivativeQuark}
\eeq
On the other hand, the gluon and ghost one-loop contributions contain 
${\cal D}_\mu^{ac}$ and ${\cal F}_{\mu\nu}^a$, which only depend on the gluon background field $\A_\mu$. This is, of course, because the gluon and ghost fields do not have electric charge and thus cannot interact with EM fields. Since these contributions are the same as in the pure Yang-Mills (YM) theory, we may call these the YM part.

So far, we have not specified the background field ${\cal A}_\mu$, but it can contain both the chromo-EM fields and the Polyakov loop. Let us briefly explain how the Polyakov loop is described within our framework. In the pure Yang-Mills theory at finite temperature, there is a confinement-deconfinement transition whose order parameter is given by the Polyakov loop. It is defined by the (closed) Wilson line along the imaginary time ($\tau$) direction:
\beq
\Phi (\vec{x})
&=& \frac{1}{N_{c}} {\rm{Tr}} \ \mathcal{P} \ {\rm{exp}} \left\{ ig \int^{\beta}_{0} d\tau {A}_{4}^{a} (\tau, \vec{x}) T^{a} \right\}\, ,
\label{defPolyakovLoop}
\eeq
where $\beta = 1/T$ is the inverse temperature and ${\cal P}$ stands for a path-ordered product along the imaginary time direction. Indeed, $\langle \Phi \rangle \to 0$ ($\langle \Phi \rangle \neq 0$) corresponds to a confining (deconfined) phase, since the negative logarithm of the expectation value of the Polyakov loop can be identified with the free energy of a static quark (a vanishing value of the Polyakov loop implies that the energy of a single quark state is infinity). These two phases are distinguished by the center symmetry. The gauge fields at finite temperature are not necessarily periodic in the direction of imaginary time and can have ambiguity related to the center subgroup $Z_{N_c}$ of the gauge symmetry SU$(N_c)$. This residual symmetry is called the center symmetry and the theory is invariant under gauge transformations which differ at $\tau = 0$ and $\tau =  \beta$ by a center element of the gauge group. The Polyakov loop $\Phi$ transforms as $\Phi \to \e^{2\pi i n/N_{c}}\Phi$ $(n=0,1,2, \ldots, N_{c}-1)$. Thus, the values of $\Phi$ distinguish the center symmetric (confining) phase and the center broken (deconfined) phase. Dynamical quarks, however, explicitly break the center symmetry. Therefore, in QCD, the Polyakov loop should be understood as an approximated order parameter. Still, we can compute an effective action for the Polyakov loop and discuss how a phase transition occurs when external parameters such as temperature are varied. 

An effective action for the Polyakov loop in the pure Yang-Mills theory was obtained in Refs.~\cite{Gross:1980br, Weiss:1980rj} in the following way:
Working in what we now call the ``Polyakov gauge" for a time-independent field $A_4^a(\vec x)=\phi(\vec x)\delta^{a3}$ in the SU(2) case, 
the authors of Refs.~\cite{Gross:1980br, Weiss:1980rj} performed a functional integral with respect to fluctuations around the field $\phi(\vec x)$. 
This procedure is nothing but the one we explained above where we treated the gluon field $A_\mu^a$ as a background ${\cal A}_\mu^a$ with a fluctuation around it. Besides, as long as we consider a spatially homogeneous and time-independent order parameter $\bar{\A}_4^a$, we can have both the Polyakov loop and the chromo-EM fields at the same time. We divide the background field into the constant part and the coordinate-dependent part as 
$\mathcal{A}_{\mu}^{a}(x) = (\bar{\A}_{\mu} + \hat{\A}_{\mu}(x)) n^{a}$.
The second term gives the real (physical) chromo-EM fields so that $\F_{\mu \nu}^{a} = \partial_{\mu} \A_{\nu}^{a}(x) - \partial_{\nu} \A_{\mu}^{a}(x) = (\partial_{\mu} \hat{\A}_{\nu}(x) - \partial_{\nu} \hat{\A}_{\mu}(x)) n^{a}$, while the first constant term $\bar{\A}_{\mu}$ does not. 
We want to treat both the chromo-EM fields and the Polyakov loop, and the latter is described at finite temperature. In order to have the both, we specify the transformation of the temporal component of the background field $\A_0^a(x)$ under the Wick rotation of the coordinate, $x_{0} \to -ix_{4} = -i\tau$ and $x_{i} \to x_{i} \ (i=1,2,3)$, as follows:
$\mathcal{A}_{0}^{a}(x) = (\bar{\A}_{0} + \hat{\A}_{0}(x)) n^{a} \to (i\bar{\A}_{4} + \hat{\A}_{0}(x)) n^{a}$.
In this way, the first term gives the Polyakov loop defined in Eq.~(\ref{defPolyakovLoop}), while the second term remains unchanged to give the real chromo-EM fields.
We work in the Polyakov gauge for $\bar{\A}_{4}^{a}$ \cite{Weiss:1980rj}\footnote{In the literature, the fourth component of the gauge field $\bar{\A}^{a}_{4}$ in the Polyakov gauge is often expressed in terms of $N_{c}-1$ real scalar fields. In our formalism, these fields are properly encoded in the color eigenvalues $\omega_{i} \ (i=1, \ldots, N_{c})$ and $v_{h} \ (h=1, \ldots, N_{c}^{2} -1)$, which will be defined later. Here, choosing the third 
direction of the color unit vector---$n^{a} = \delta^{a 3}$ at finite temperature---we pick up the one particular field $\bar{\A}_{4}$ which provides a simple expression for the Poyakov loop as shown in Eq.~(\ref{simple_Polyakov_loop}). However, in the finial expression of our effective action, it is quite straightforward to keep all the $N_{c}-1$ scalar fields in the color eigenvalues $\omega_{i}$ and $v_{h}$.}:
\beq
\bar{\A}_{4}^{a} = \bar{\A}_{4}\, \delta^{3 a}, \ \ \ \partial_{4} \bar{\A}_{4} = 0\, , 
\eeq
which does not conflict with the covariantly constant condition in Eq.~(\ref{CondtionCovariantConstant}). Notice that we use this gauge with $\delta^{a3}$ even for the SU($N_c$) case, and the color unit vector $n^a$ introduced in Eq.~(\ref{BackgroundField}) should be understood as $n^a=\delta^{3a}$ at finite temperature.\footnote{Still, we keep the expression $n^a$ because we will discuss the case at zero temperature.} Following Ref.~\cite{Weiss:1980rj}, we also introduce a dimensionless field $C$ as 
\beq
C = \frac{g {\bar{\A}_{4} } }{ 2 \pi T }, 
\eeq
so that the Polyakov loop is simply given as 
\beq
\Phi
&=& {\rm{cos}} (\pi C) \qquad \qquad \quad \ \ {\rm{for \ SU(2) }}\, ,  \nonumber \\
\Phi
&=& \frac{1}{3} \Big\{ 1 + 2{\rm{cos}}( \pi C) \Big\} \quad \ {\rm{for\ SU(3)}}\,  . \label{simple_Polyakov_loop}
\eeq

\subsection{Yang-Mills part of effective action}

Now, we consider the Yang-Mills part (gluon and ghost contributions) of the one-loop effective action. In the one-loop level, the effect of EM fields is not included in gluon and ghost loops, since these do not directly interact with EM fields. From Eq.~(\ref{full_actions}), the effective actions of gluon and ghost parts are given, respectively, as
\beq
iS_{\rm gluon}
&\equiv & {\rm{ln}} \ {\rm{det}} \left[ - ({\cal D}^{2})^{ac} g_{\mu \nu} - 2 g f^{abc} {\cal F}^{b}_{\mu \nu} \right]^{-\frac12}, \label{gluon-part}\\
iS_{\rm ghost}
&\equiv & {\rm{ln}} \ {\rm{det}} \left[ - ({\cal D}^{2})^{ac}  \right]^{+1}.
\label{ghost-part}
\eeq
Let us first explore the gluon part (\ref{gluon-part}). By using the proper time integral,\footnote{We use the following identity:
$$
\ln (\hat M -i\delta)=\frac{1}{\epsilon}- \frac{i^\epsilon}{\epsilon \Gamma(\epsilon)}\int_0^\infty \frac{ds}{s^{1-\epsilon}}\, \e^{-is (\hat M -i\delta)} $$ 
in the limit $\epsilon\to 0$ and $\delta\to 0$. We ignore the first divergent term, since it does not depend on the fields.} the gluon part of the effective action can be rewritten in the following form (the limit $\epsilon,\delta\to 0$ is always implicit and should be taken after the calculation): 
\beq
iS_{\rm gluon}
&=& -\frac{1}{2} {\rm{Tr}} \ {\rm{ln}} 
\left[ - ({\cal D}^{2})^{ac} g_{\mu \nu} 
       - 2 g f^{abc} {\cal F}^{b}_{\mu \nu} 
\right] \nonumber \\
&=& \int d^{4}x \frac{i^{\epsilon}}{2} 
      \sum_{h=1}^{N_{c}^{2}-1} \int^{\infty}_{0} \frac{ds}{s^{1-\epsilon}} 
      {\rm{tr}} \langle x | 
        \e^{- i \left( - {\cal D}_{v_{h}}^{2} g_{\mu \nu} 
                  + 2i g v_{h} {\cal F}_{\mu \nu} -i \delta \right) s}
                | x \rangle \nonumber \\
&=& \int d^{4}x  \frac{i^{\epsilon}}{2} \sum_{h=1}^{N_{c}^{2}-1} 
       \int^{\infty}_{0} \frac{ds}{s^{1-\epsilon}} 
        \e^{-\delta s} 
          \left\{ \e^{-i(2gv_{h}\aaa )s} + \e^{ -i ( - 2gv_{h}\aaa )s} 
                + \e^{-i ( igv_{h}\bbb )s } + \e^{ -i ( -2igv_{h}\bbb )s } 
          \right\} \nonumber \\
&& \times \langle x | \e^{ - i ( - {\cal D}_{v_{h}}^{2} ) s} |x \rangle \, .
\eeq
While the capital trace ``Tr" in the first line is taken with respect to colors, Lorentz indices, and coordinates, ``tr" in the second line is only for Lorentz indices. Also, in the second line, we have introduced real quantities $v_{h}$ $(h=1,\ldots, N_c^2-1)$ that are eigenvalues of a Hermitian matrix $V^{ac}\equiv if^{abc} {n}^{b}$ (i.e., $V^{ac}\varphi^c=v_h \varphi^a$), and Lorentz-invariant quantities
$\aaa$, $\bbb$ defined by
\beq
\aaa
\equiv \frac{1}{2} \sqrt{ \sqrt{ \F^{4} + (\F\cdot \tilde{\F})^{2} } + \F^{2} }\, , \ \ \ \ 
\bbb
\equiv \frac{1}{2} \sqrt{ \sqrt{ \F^{4} + (\F\cdot \tilde{\F})^{2} } - \F^{2} }\, ,
\eeq
with the dual field-strength tensor $\tilde{\F}^{\mu \nu} = \frac{1}{2} \epsilon^{\mu \nu \alpha \beta} \F_{\alpha \beta}$ (or equivalently, by $\aaa^2-\bbb^2=\frac12 \F^2$ and $\aaa \bbb = \frac14 \F\cdot \tilde \F$).
The covariant derivative is defined as ${\cal D}_{v_{h} \mu} = \partial_{\mu} - ig v_{h} \A_{\mu}$. 
The calculation up to now is in fact the same as in the case at zero temperature which was done in Ref.~\cite{Ozaki:2013sfa}. 
At finite temperature, however, one needs to be careful in evaluating the matrix element $\langle x | \e^{ - i ( - {\cal D}_{v_{h}}^{2} ) s} |x \rangle$.
Namely, it can be now written as the Matsubara summation:
\beq
\langle x | \e^{ - i ( - {\cal D}_{v_{h}}^{2} ) s} |x \rangle
&=& \left. i T \sum_{n=-\infty}^{\infty} \int \frac{ d^{3}p }{ (2\pi)^{3} } 
\, \e^{- p_\alpha X_{h}^{\alpha\beta}(is) p_\beta }\, \e^{-Y_{h}(is) } \right|_{p_{0} = igv_{h}\bar{\A}_{4} - i 2\pi n T}\, ,
\label{Matrix-element-T} 
\eeq
where the functions $X_{h}^{\alpha \beta}(\bar{s})$ and $Y_{h}(\bar{s})$ have been defined as \cite{Dittrich:2000zu}
\beq
X_{h}^{\alpha \beta}(\bar{s})
&=& \left[(gv_{h}\F )^{-1} {\rm{tan}}(gv_{h}\F \bar{s})\right]^{\alpha\beta}, \nonumber \\
Y_{h}(\bar{s})
&=& \frac{1}{2} {\rm{tr}} \ {\rm{ln}} \ {\rm{cos}}(gv_{h}\F \bar{s}).
\eeq
In the presence of the Polyakov loop $\bar{\A}_{4}$, the periodic boundary condition of the gluon in the imaginary time direction is modified.
Then, the Matsubara frequency is shifted by the Polyakov loop as in Eq.~(\ref{Matrix-element-T}).
Performing the three-dimensional momentum integral and applying the Poisson resummation~\cite{Dittrich:2000zu}, one can obtain the matrix element in terms of 
$\aaa$ and $\bbb$ as
\beq
\!\!\langle x | \e^{ - i ( - {\cal D}_{v_{h}}^{2} ) s} |x \rangle
&=& -\frac{i}{16 \pi^{2}} \frac{ gv_{h}\aaa s}{{\rm{sin}}(gv_{h}\aaa s)} \frac{gv_{h}\bbb s}{ {\rm{sinh}}(gv_{h}\bbb s) } \left[
1 + 2\sum_{n=1}^{\infty} \e^{ i \frac{\mathfrak{h}(s)}{4T^{2}}n^{2}} {\rm{cos}}\left( \frac{ gv_{h} \bar{\A}_{4}}{T} n \right) \right],
\label{KarnelG}
\eeq
where
\beq
\mathfrak{h}(s)
&=& \frac{\bbb^{2} - {\eee}^{2} }{\aaa^{2}+\bbb^{2}}\, g v_{h} \aaa\, {\rm{cot}}(gv_{h} \aaa s) + \frac{ \aaa^{2} + {\eee}^{2} }{ \aaa^{2} + \bbb^{2} }\, g v_{h} \bbb\, {\rm{coth}} (gv_{h}\bbb s)\, ,
\eeq
with
\beq  
{\eee}^{2} = (u_{\alpha} \F^{\alpha \mu})( u_{\beta} \F^{\beta}_{\mu} ).
\eeq
The vector $u^{\mu}$ is the heat-bath four-vector, which is $(1,0,0,0)$ in the rest frame of the heat bath. The first (second) term in Eq.~(\ref{KarnelG}) corresponds to the zero-(finite-)temperature contribution. 
The gluon part of the effective action is then given as
\beq
iS_{\rm gluon} 
&=& - \frac{i^{1+\epsilon}}{ 32 \pi^{2} } \int d^{4}x \sum_{h=1}^{N_{c}^{2}-1} 
\int^{\infty}_{0} \frac{ds}{s^{3-\epsilon}} \e^{-\delta s} 
\left\{ \e^{-i(2gv_{h}\aaa )s} + \e^{ -i ( - 2gv_{h}\aaa )s} 
      + \e^{-i ( igv_{h}\bbb )s } + \e^{ -i ( -2igv_{h}\bbb )s } 
\right\} \nonumber \\
&&\qquad\qquad\ \ \times \frac{ gv_{h}\aaa s}{{\rm{sin}}(gv_{h}\aaa s)} 
\frac{gv_{h}\bbb s}{ {\rm{sinh}}(gv_{h}\bbb s) } \left[
1 + 2\sum_{n=1}^{\infty} \e^{ i \frac{\mathfrak{h}(s)}{4T^{2}}n^{2}} 
{\rm{cos}}\left( \frac{ gv_{h} \bar{\A}_{4}}{T} n \right) \right].\label{action_gluon}
\eeq
Similarly, we obtain the ghost part as
\beq
iS_{\rm ghost}
&=& \frac{i^{1+\epsilon}}{ 32 \pi^{2} } \int d^{4}x \sum_{h=1}^{N_{c}^{2}-1} 
\int^{\infty}_{0} \frac{ds}{s^{3-\epsilon}} \e^{-\delta s} 
\left\{ 2 \right\} \nonumber \\
&& \times \frac{ gv_{h}\aaa s}{{\rm{sin}}(gv_{h}\aaa s)} 
\frac{gv_{h}\bbb s}{ {\rm{sinh}}(gv_{h}\bbb s) } \left[
1 + 2\sum_{n=1}^{\infty} \e^{ i \frac{\mathfrak{h}(s)}{4T^{2}}n^{2}} 
{\rm{cos}}\left( \frac{ gv_{h} \bar{\A}_{4}}{T} n \right) \right].
\label{action_ghost}
\eeq
In both parts, the first terms in the square brackets are the results at zero temperature and agree with the known results \cite{Ozaki:2013sfa}. As discussed in detail in Ref.~\cite{Ozaki:2013sfa}, each term has an ultraviolet (UV) divergence, which, however, can be absorbed by renormalizing the coupling $g$ and fields $\A_\mu$ \cite{Savvidy:1977as, Matinyan:1976mp}. On the other hand, the finite-temperature contributions do not have UV divergence, and thus we do not need an additional renormalization procedure for the finite-temperature contributions. 
We regard the coupling and fields as renormalized ones and focus on UV-finite pieces in Eqs.~(\ref{action_gluon}) and (\ref{action_ghost}).

Our results (\ref{action_gluon}) and (\ref{action_ghost}) are effective actions for chromo-EM fields as well as the Polyakov loop at finite temperature. These are generalizations of the previous results in two cases. Indeed, if we consider the pure chromo-{\it electric} background with a Polyakov loop (${\cal B}=0,\ {\cal E}\neq 0, \ \A_0\neq 0$), we find $\aaa \to i{\cal E}$, $\bbb \to 0$ and reproduce Gies's effective action at finite temperature \cite{Gies:2000dw}. Moreover, in the case of the pure chromo-{\it magnetic} background (${\cal E}=0,\ {\cal B}\neq 0$, $\bar{\A}_{4}=0$), we find $\aaa \to {\cal B}$, $\bbb \to 0$ and reproduce the results obtained in Refs.~\cite{Dittrich:1980nh, Kapusta:1981nf}.

\subsection{Quark part of effective action}

For the quark part of the effective action, we follow basically the same procedures as in the Yang-Mills part.
From the functional integral (\ref{full_actions}), the quark part of the one-loop effective action reads
\beq
iS_{\rm quark}
&=& {\rm{ln}} \ {\rm{det}} \left[ i \gamma_{\mu} \hat{\cal D}^{\mu} - M_{q} \right].
\eeq
Utilizing the proper time integral, we evaluate the effective action as
\beq
iS_{\rm quark}
&=& {\rm{Tr}} \ {\rm{ln}} \left[ i \gamma_{\mu} \hat{\cal D}^{\mu} - M_{q} \right] \nonumber \\
&=& -\int dx^{4} \frac{i^{\epsilon}}{2} \sum_{i=1}^{N_{c}} \sum_{f=1}^{N_{f}} \int^{\infty}_{0} \frac{ds}{s^{1-\epsilon}} 
\e^{-i (m_{q_{f}}^{2} - i \delta ) s} {\rm{tr}} \langle x | 
\e^{-is \left( -\mathbb{D}_{i,f}^{2} - \frac{1}{2} \sigma \cdot \mathbb{F}_{i,f} \right) } 
|x \rangle,
\eeq
where $\mathbb{D}_{i,f}^{\mu} = \partial^{\mu} - i \mathbb{A}_{i,f}^{\mu}$ with the field $\mathbb{A}_{i,f}^{\mu}$ being a linear combination of the gluon field $\A_{\mu}$ and the photon field $a^{\mu}$ as 
\beq
\mathbb{A}_{i,f}^{\mu}
&=& g\omega_{i} \A^{\mu} + eQ_{q_{f}} a^{\mu}. \label{linear_combination}
\eeq
This covariant derivative $\mathbb{D}_{i,f}^{\mu}$ can be obtained from $\hat{\cal D}^{\mu}$ defined in Eq. (\ref{CovariantDerivativeQuark}) with the covariantly constant field employed as the background field.
Here $\omega_{i}\ (i=1,\ldots,N_c)$ are eigenvalues of an $N_c\times N_c$ matrix ${n}^{a} T^{a}$ and satisfy\footnote{Let $\Omega$ be a diagonal matrix with eigenvalues $\omega_i$, i.e., $\Omega={\rm diag}(\omega_1,\ldots,\omega_{N_c})=Un^aT^aU^\dagger$. Then, $\sum_{i=1}^{N_c}\omega_i=\tr \Omega= n^a\tr T^a=0$ and $\sum_{i=1}^{N_c}\omega_i^2=\tr \Omega^2=\tr(T^aT^b)n^a n^b=1/2.$ } $\sum_{i=1}^{N_c}\omega_i=0$ and $\sum_{i=1}^{N_c}\omega_i^2=1/2$.
The field-strength tensor $\mathbb{F}_{i,f}^{\mu \nu}$ can be expressed in terms of constant chromo-EM fields $\vec{\mathcal{E}}$, $\vec{\mathcal B}$, and EM fields $\vec{E}$, $\vec{B}$ as [with the notation $\vec{V}=(V_x,V_y,V_z)$]
\beq
\mathbb{F}_{i,f}^{\mu \nu}
&=& g \omega_{i} {\cal F}^{\mu \nu} + eQ_{q_{f}} f^{\mu \nu} \nonumber \\
&=&
g \omega_{i} \left(
\begin{array}{cccc}
0 & \mathcal{E}_{x} & \mathcal{E}_{y} & \mathcal{E}_{ z } \\
-\mathcal{E}_{x} & 0 & \mathcal{B}_{z} & - \mathcal{B}_{y} \\
-\mathcal{E}_{y} & - \mathcal{B}_{z} & 0 & \mathcal{B}_{x} \\
-\mathcal{E}_{z} & \mathcal{B}_{y} & - \mathcal{B}_{x} & 0 \\
\end{array}
\right)
+
e Q_{q_{f}}  \left(
\begin{array}{cccc}
0 & E_{x} & E_{y} & E_{z } \\
-E_{x} & 0 & B_{z} & - B_{y} \\
-E_{y} & - B_{z} & 0 & B_{x} \\
-E_{z} & B_{y} & - B_{x} & 0 \\
\end{array}
\right).
\label{matrix_field}
\eeq
The eigenvalues of the field-strength tensor $\mathbb{F}^{\mu \nu}_{i,f}$ are given by $\pm i\aaa_{i,f}$ and $\pm \bbb_{i,f}$
with
\beq
\aaa_{i,f}
 = \frac{1}{2} \sqrt{  \sqrt{ \mathbb{F}_{i,f}^{4} + ( \mathbb{F}_{i,f} \cdot \tilde{\mathbb{F}}_{i,f} )^{2} } + \mathbb{F}_{i,f}^{2} } \ , \ \ \ \ \ 
\bbb_{i,f}
 = \frac{1}{2} \sqrt{  \sqrt{ \mathbb{F}_{i,f}^{4} + ( \mathbb{F}_{i,f} \cdot \tilde{\mathbb{F}}_{i,f} )^{2} } - \mathbb{F}_{i,f}^{2} } \ .
\eeq
The dual field-strength tensor $\tilde{\mathbb{F}}^{\mu \nu}_{i,f}$ is defined as $\tilde{\mathbb{F}}^{\mu \nu}_{i,f}  = \frac{1}{2} \epsilon^{\mu \nu \alpha \beta} \mathbb{F}_{i,f \alpha \beta}$. By using Eq.~(\ref{matrix_field}), $\mathbb{F}_{i,f}^{2}=2(\aaa_{i,f}^2-\bbb_{i,f}^2)$ and $\mathbb{F}_{i,f} \cdot \tilde{\mathbb{F}}_{i,f}=4\aaa_{i,f}\bbb_{i,f}$ can be expressed in terms of chromo-EM fields and EM fields as
\beq
\mathbb{F}_{i,f}^{2}
&=& 2( \vec{ \mathcal{B} }_{i,f}^{2} - \vec{ \mathcal{E} }_{i,f}^{2} ), \nonumber \\
\mathbb{F}_{i,f} \cdot \tilde{\mathbb{F}}_{i,f}
&=& -4 \vec{ \mathcal{E} }_{i,f} \cdot \vec{ \mathcal{B} }_{i,f},
\label{FFtilde}
\eeq
where we have defined the combined electromagnetic fields as $\vec{ \mathcal{E} }_{i,f} = g \omega_{i} \vec{ \mathcal{E} } + eQ_{q_{f}} \vec{E}$ and $\vec{ \mathcal{B} }_{i,f} = g \omega_{i} \vec{ \mathcal{B} } + eQ_{q_{f}} \vec{ B}$. 
Taking the trace of the matrix $\langle x | \e^{-is \left( -\mathbb{D}_{i,f}^{2} - \frac{1}{2} \sigma \cdot \mathbb{F}_{i,f} \right) } |x \rangle$ at finite temperature, we get
\beq
&&{\rm{tr}} \langle x | \e^{-is \left( -\mathbb{D}_{i,f}^{2} - \frac{1}{2} \sigma \cdot \mathbb{F}_{i,f} \right) } |x \rangle \nonumber \\
&&\qquad = \left. i T  \sum_{n=-\infty}^{\infty} \int \frac{ d^{3} p }{ (2\pi)^{3} }\,  \e^{ - p_\alpha \mathbb{X}^{\alpha\beta}_{i,f}(is) p_\beta } \e^{- \mathbb{Y}_{i,f}(is) }\, {\rm{tr}} \, \e^{  \frac{i}{2} \sigma \cdot \mathbb{F}_{i,f}s } \right|_{p_{0} = ig \omega_{i} \bar{\A}_{4} - i \pi (2n+1) T}\, .
\label{matrix_quark}
\eeq
Here, the functions $\mathbb{X}_{i,f}^{\alpha \beta} (\bar{s})$ and $\mathbb{Y}_{i,f} (\bar{s})$ have been defined as~\cite{Dittrich:2000zu}
\beq
\mathbb{X}_{i,f}^{\alpha \beta} (\bar{s})
&=& \left[ \mathbb{F}_{i,f}^{-1}\, {\rm{tan}} ( \mathbb{F}_{i,f} \bar{s} ) \right]^{\alpha \beta}, \nonumber \\
\mathbb{Y}_{i,f} (\bar{s})
&=& \frac{1}{2} {\rm{tr}} \ {\rm{ln}} \ {\rm{cos}} ( \mathbb{F}_{i,f} \bar{s} )\, .
\eeq 
In the presence of the Polyakov loop $\bar{\A}_{4}$, the antiperiodic boundary condition for the quark is also modified. Then, the temporal component of the four-momentum vector has been replaced by the Polyakov loop and the Matsubara frequency for a fermion in Eq.~(\ref{matrix_quark}). 
The third part, ${\rm{tr}}\, \e^{  \frac{i}{2} \sigma \cdot \mathbb{F}_{i,f} s}$, is common with the case at zero temperature and was computed in Ref.~\cite{Ozaki:2013sfa}. The result is
\beq
{\rm{tr}} \, {\rm{exp}} \left( \frac{i}{2} \sigma \cdot \mathbb{F}_{i,f} s \right)
&=& 4 {\rm{cos}}( \aaa_{i,f}s ) {\rm{cosh}} (\bbb_{i,f} s).
\eeq
Now, performing the three-dimensional momentum integral and using the Poisson resummation, we find from Eq.~(\ref{matrix_quark})
\beq
{\rm{tr}} \langle x | \e^{-is \left( -\mathbb{D}_{i,f}^{2} - \frac{1}{2} \sigma \cdot \mathbb{F}_{i,f} \right) } |x \rangle 
&=&  - \frac{ i }{ 4 \pi^{2} s^{2} } \frac{ ( \aaa_{i,f}s )( \bbb_{i,f}s ) }{ {\rm{sin}}( \aaa_{i,f} s ) {\rm{sinh}} ( \bbb_{i,f} s ) } {\rm{cos}}( \aaa_{i,f}s ) {\rm{cosh}}( \bbb_{i,f}s ) \nonumber \\
&& \times \left\{ 1 + 2 \sum_{n=1}^{\infty} (-1)^{n} \e^{ \frac{i}{4T^{2}} \mathfrak{h}_{i,f}(s) n^{2} } {\rm{cos}} \left( \frac{ g \omega_{i}\bar{\A}_{4} n }{ T} \right) \right\} \, ,
\eeq
where
\beq
\mathfrak{h}_{i,f}(s)
&=& \frac{ \bbb_{i,f}^{2} - {\eee}_{i,f}^{2} }{\aaa_{i,f}^{2} + \bbb_{i,f}^{2}}\aaa_{i,f} {\rm{cot}}(\aaa_{i,f}s) + \frac{ \aaa_{i,f}^{2} + {\eee}_{i,f}^{2} }{ \aaa_{i,f}^{2} + \bbb_{i,f}^{2} } \bbb_{i,f} {\rm{coth}}(\bbb_{i,f}s)\, ,
\eeq
with
\beq
{\eee}_{i,f}^{2}
&=& (u_{\alpha} \mathbb{F}_{i,f}^{\alpha \mu}) ( u_{\beta} \mathbb{F}^{\ \beta}_{i,f \mu})\, .
\eeq
In the heat-bath rest frame, we have $u^{\mu} = (1,0,0,0)$ and then
$
{\eee}_{i,f}^{2}  = \vec{ \mathcal{E} }_{i,f}^{2} = ( g \omega_{i} \vec{\mathcal{E}} + eQ_{q_{f}} \vec{E} )^{2}
$.
Therefore, the quark part of the one-loop effective action reads
\beq 
iS_{\rm quark}
&=& \frac{i^{1+\epsilon}}{8\pi^{2}} \int d^{4}x \sum_{i=1}^{N_{c}} \sum_{f=1}^{N_{f}} \int_{0}^{\infty} \frac{ds}{s^{3-\epsilon}} 
\e^{-i(m_{q_f}^{2}-i\delta)s} (\aaa_{i,f}s)(\bbb_{i,f}s) {\rm{cot}}(\aaa_{i,f}s) {\rm{coth}}(\bbb_{i,f}s) \nonumber \\
&&\qquad \times \left[ 1 + 2 \sum_{n=1}^{\infty} (-1)^{n} \e^{\frac{i}{4T^{2}} \mathfrak{h}_{i,f}(s) n^{2} } {\rm{cos}}\left( \frac{ g\omega_{i} \bar{\A}_{4}n}{T} \right) \right]\, .
\label{action_quark}
\eeq
As in the YM part, the first (second) term corresponds to the zero-(finite-)temperature contribution. The zero-temperature contribution agrees with the previous result obtained in Ref.~\cite{Ozaki:2013sfa}. 

Again, the first term contains UV divergences. 
These divergences have two origins: QCD and QED \cite{Ozaki:2013sfa}. 
This is because the resummed quark one-loop diagrams contain contributions from the diagrams with only two EM field insertions (QED) and only two chromo-EM field insertions (QCD).
The UV divergence coming from purely QCD dynamics is additive to the one which we encounter in the YM part. Then, we can absorb all the UV divergences by renormalizing the coupling $g,e$ and fields $\mathcal{A}_{\mu}, a_{\mu}$. 
From the renormalization procedure at zero temperature, we have obtained the correct beta functions of both QCD and QED in Ref.~\cite{Ozaki:2013sfa}. The sum of the three parts (\ref{action_gluon}), (\ref{action_ghost}), and (\ref{action_quark}) may be called the Euler-Heisenberg-Weiss action in QCD+QED at finite temperature. This result can be applied to several systems where strong EM fields and chromo-EM fields coexist at zero and finite temperatures. In the next section, we will show some applications of our effective actions.\\

\section{Applications of Euler-Heisenberg-Weiss action in QCD+QED}

In this section we will discuss two applications of our results. 
The first one is the quark pair production in the presence of both EM and chromo-EM fields. We treat the effective action at zero temperature. The second application is to investigate the effects of EM fields on the effective potential for the Polyakov loop at finite temperature. We will discuss the possible implication for the inverse magnetic catalysis.

\subsection{Quark pair production in QCD+QED fields}

Let us first discuss quark-antiquark pair production in constant QCD+QED fields as an application of our effective action. For this problem, only the quark part (\ref{action_quark}) is relevant.

In the early stage of relativistic heavy-ion collisions, extremely strong chromo-EM fields and EM fields could coexist. 
Notice that the strong {\it{electric}} field in addition to the strong magnetic field could be created on an event-by-event basis~\cite{Deng:2012pc}.
The strength of the chromo-EM fields is approximately of the order of the saturation scale: $|g\vec{\mathcal{B}}|, |g\vec{\mathcal{E}}| \sim Q_{s}^2 $, whereas strengths of EM fields would reach the QCD nonperturbative scale $|e\vec E|, |e\vec B|\sim \Lambda_{QCD}^2$, or even exceed it. Under such strong QCD+QED fields, a number of quark-antiquark pairs must be created through the Schwinger mechanism. The pair-production rate per unit space-time volume can be obtained from the imaginary part of the quark effective Lagrangian at zero-temperature. 
Taking the zero temperature contribution in Eq.~(\ref{action_quark}), one finds
\beq
\mathcal{L}_{\rm quark}
= \frac{ S_{\rm quark} }{ \int d^{4}x } 
= \frac{1}{8\pi^{2}} \sum_{i=1}^{N_{c}} \sum_{f=1}^{N_{f}} \int^{\infty}_{0} \frac{ds}{s^{3}} \e^{-is(m_{q_{f}}^{2} - i \delta) } (\aaa_{i,f}s)(\bbb_{i,f}s) {\rm{cot}}(\aaa_{i,f}s) {\rm{coth}}(\bbb_{i,f}s)\, .
\eeq
This is the same as the result obtained in Ref.~\cite{Ozaki:2013sfa}. 
The imaginary part of the effective Lagrangian thus reads
\beq
{\Im }m\, \mathcal{L}_{\rm quark}
&=& - \frac{1}{8\pi^{2}} \sum_{i=1}^{N_{c}} \sum_{f=1}^{N_{f}} \int^{\infty}_{0} \frac{ds}{s^{3}} \e^{- \delta s } {\rm{sin}}(m_{q_{f}}^{2} s) \times (\aaa_{i,f}s)(\bbb_{i,f}s) {\rm{cot}}(\aaa_{i,f}s) {\rm{coth}}(\bbb_{i,f}s) \nonumber \\
&=& \frac{1}{2 i } \frac{1}{8 \pi^{2} } \sum_{i=1}^{N_{c}^{2}} \sum_{f=1}^{N_{f}} \left\{
\int^{0}_{-\infty} \frac{ds}{s^{3}}\, \e^{-is(m_{q_{f}}^{2} + i\delta ) } + \int^{\infty}_{0} \frac{ds}{s^{3}}\, \e^{-is(m_{q_{f}}^{2} - i \delta ) } \right\} \nonumber \\
&& \qquad \times (\aaa_{i,f}s)(\bbb_{i,f}s) {\rm{cot}}(\aaa_{i,f}s) {\rm{coth}}(\bbb_{i,f}s)\, .
\eeq
The integrand has infinitely many poles along the real axis [from cot$( \aaa_{i,f}s)$] and along the imaginary axis [from coth$(\bbb_{i,f}s)$]. With a small positive number $\delta>0$, the integral contour along the real axis is inclined.  Closing the contour in the lower half of the $s$ plane as depicted in Fig. 2 and picking up the poles lying on the imaginary axis $s_{\rm poles} = - i n \pi / \bbb_{i,f}$, we find
\begin{figure}[t]
\begin{minipage}{0.8\hsize}
\begin{center}
\includegraphics[width=0.8 \textwidth]{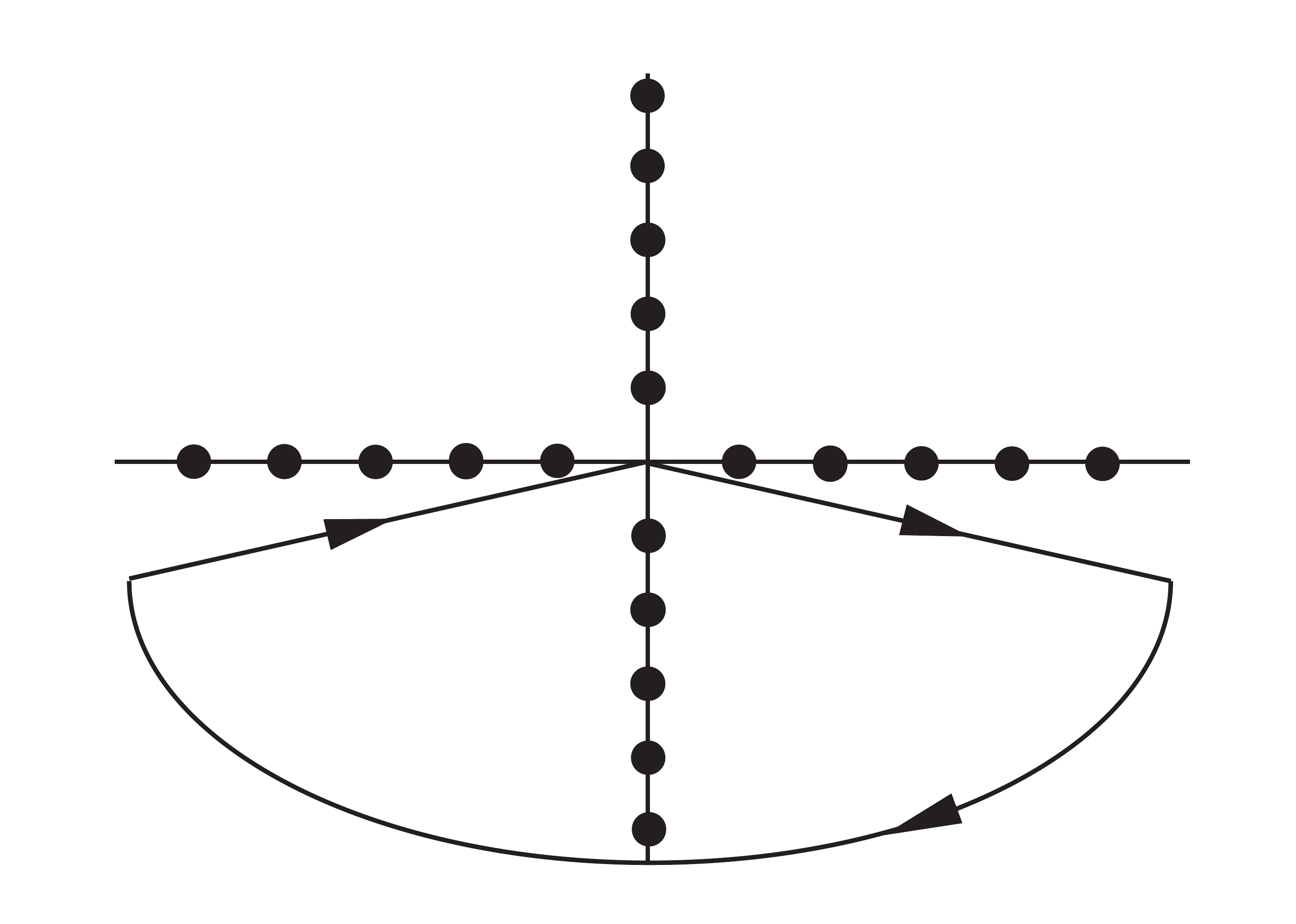}
\vskip -0.1in
\end{center}
\end{minipage}
\caption{
Contour on the complex $s$ plane. The contour along the real axis is inclined by an infinitesimal number $\delta>0$.}
\end{figure}
\beq
{\Im}m\, \mathcal{L}_{\rm quark}
&=& \frac{1}{8 \pi^{2}} \sum_{i=1}^{N_{c}} \sum_{f=1}^{N_{f}} \aaa_{i,f} \bbb_{i,f} \sum_{n=1}^{\infty} \frac{1}{n}\, \e^{ - \frac{ m_{q_{f}}^{2} }{\bbb_{i,f}} n\pi }
 {\rm{coth}} \left( \frac{\aaa_{i,f}}{\bbb_{i,f}} n \pi \right).
\label{ImLq_full}
\eeq
By using this expression, we can investigate quark-antiquark pair productions under arbitrary configurations of constant chromo-EM and EM fields. 
The production rate per unit space-time volume is given by 
$
w_{q\bar{q}} = 2{\Im}m\, \mathcal{L}_{\rm quark}.
$
When we take $N_{c} = N_{f}=1$, $Q=1$, $g\to 0$, $B\to 0$ and replace $m_{q} \to m_{e}$ in Eq.~(\ref{ImLq_full}), we reproduce the well-known Schwinger formula for the production rate of $e^+e^-$ pairs in an electric field \cite{Schwinger:1951nm}:
\beq
w_{e^{+}e^{-}} = 2{\Im m}\, \mathcal{L}_{\rm EH} 
= \frac{(eE)^{2}}{4 \pi^{3}}   \sum_{n=1}^{\infty} \frac{1}{n^{2}} \e^{ - \frac{ m_{e}^{2} }{eE} n\pi },
\label{rate_e}
\eeq
as we expected.
On the other hand, in the pure chromo-electric field case, we obtain the same formula for quark productions derived by G.C.~Nayak \cite{Nayak:2005pf}.

\subsubsection{Quark pair production in purely electric background}

First, we shall consider quark pair production in a purely electric background with vanishing magnetic fields: $\vec{B}, \vec{\mathcal{B}} \to 0$.
In this case, the production rate for $q\bar q$ pairs of flavor $f$ becomes
\beq
w_{q_f \bar q_f}
&=& \frac{1}{4 \pi^{3}} \sum_{i=1}^{N_{c}} \bbb_{i,f}^{2} \sum_{n=1}^{\infty} \frac{1}{n^{2}}\, \e^{- \frac{ m_{q_{f}}^{2}}{\bbb_{i,f}} n \pi}, 
\label{EcEformula}
\eeq
where 
$\bbb_{i,f}= \sqrt{ \vec{ \mathcal{E} }_{i,f}^{2} } = \sqrt{ (g \omega_{i})^{2} \mathcal{E}^{2} 
                 + (eQ_{q_{f}})^{2} E^{2} 
                 + 2g \omega_{i}eQ_{q_{f}} \mathcal{E} E 
                      {\rm{cos}}\theta_{\mathcal{E}E} 
                 }$,
with 
$E = \sqrt{ \vec{E}^{2} }$, $\mathcal{E} = \sqrt{ \vec{ \mathcal{E} }^{2} }$,
and $\theta_{\mathcal{E}E}$ being the angle between $\vec{E}$ and $\vec{\mathcal{E}}$. 
For $N_{c} = 3$, the eigenvalues $\omega_{i}$ are given by $\omega_{1} = 1/2$, $\omega_{2} = -1/2$, and $\omega_{3}=0$. 
Recall that a factor $g\omega_i$ plays the role of an effective coupling between the chromo-EM field and quarks [see Eq.~(\ref{linear_combination})]. Thus, a quark (or an antiquark) with $\omega_3=0$ does not interact with the chromo-EM field in this representation. Still, since there is always a coupling with the EM fields, $q\bar q$ production with $\omega_3=0$ is possible due to electric fields, i.e., $\bbb_{i=3,f}=|eQ_{q_f} E|\neq 0$. 

Let us see the dependences of production rates on the quark mass $m_q$ and the angle $\theta_{\mathcal{E}E}$. We first consider the case with light quark masses $m_{q_f}^2\ll \bbb_{i,f}$. 
The left panel of Fig. 3 shows the light (up) quark production rate with $m_{q}=5$ MeV and $Q_{q}=+2/3$. 
The chromo-electric field is fixed to $g\mathcal{E}=1$~GeV$^{2}$, which is a typical value realized in heavy-ion collisions at RHIC and LHC, while we take several values of strength for the $E$ field. The production rate increases with increasing $E$ field, which is an expected behavior of the usual Schwinger mechanism, but it does not show dependence on the angle $\theta_{{\cal E}E}$, while $\bbb_{i,f}$ certainly depends on $\theta_{{\cal E}E}$.
This unexpected behavior can be understood as follows:
When the quark mass is small enough, $m_{q}^2 \ll \bbb_{i,f}$, we can approximate the production rate as
\beq
w_{q_f\bar q_f}
\sim \frac{1}{4 \pi^{3}} \sum_{i=1}^{N_{c}} \bbb_{i}^{2} \sum_{n=1}^{\infty} \frac{1}{n^{2}} = \frac{1}{4\pi^{3}} \left\{ \frac{ (g\mathcal{E})^{2} }{2} + N_{c}(eQ_{q}E)^{2} \right\} \zeta(2)\, ,
\eeq
where $\zeta(2) = \pi^{2}/6$ and $\bbb_{i} = \sqrt{ (g \omega_{i})^{2} \mathcal{E}^{2} + (eQ_{q})^{2} E^{2} + 2g \omega_{i}eQ_{q} \mathcal{E} E {\rm{cos}}\theta_{\mathcal{E}E} }$.
Notice that the angle dependence in $\bbb_i$ drops out thanks to the relations $\sum_{i=1}^{N_{c}} \omega_{i}^{2} = 1/2$ and $\sum_{i=1}^{N_{c}} \omega_{i} = 0$.
Therefore, the production rate is independent of the angle $\theta_{\mathcal{E}E}$.

We next discuss the production of heavy quark-antiquark pairs. Since the heavy quark limit just implies that the pair creation does not occur, we consider the case where quark masses are comparable to the background field $m_q^2 \sim \bbb_{i,f}$. This is realized for charm quarks if we again take the typical value of the chromo-electric field $g{\cal E}=1~$GeV$^2$. For $m_{c}=1.25$~GeV and $Q_{q}= Q_{\rm charm} = +2/3$, the production rate of a charm quark pair is shown in the right panel of Fig.~3. This time, while the production rate becomes small, one can see a clear dependence on the angle $\theta_{{\cal E}E}$. Both effects (small production rate and angle dependence) come from the exponential factor in Eq.~(\ref{EcEformula}).
In particular when the electric field is parallel (or antiparallel) to the chromo-electric field, the production rate has a maximum.
Since the exponential factor is very sensitive to the change of $\bbb_{i,f}$, the rate is largely enhanced at $\theta_{{\cal E}E}= 0, \pi$. 
Symmetric shape of the angle dependence with respect to $\theta_{{\cal E}E}=\pi/2$ is not so trivial. Notice that the effective field strengths of the combined field at $\theta_{{\cal E}E}= 0$ and $\pi$ are not equivalent for a fixed value of $i$; namely, it is the strongest for the parallel configuration (for $\omega_i>0$) $\bbb_{i,{\rm charm}}(\theta_{{\cal E}E}= 0)=\sqrt{ (g \omega_{i})^{2} \mathcal{E}^{2} 
                 + (eQ_{\rm charm})^{2} E^{2} 
                 + 2g \omega_{i}eQ_{\rm charm} \mathcal{E} E 
                 }$ and the weakest for the antiparallel configuration
$\bbb_{i,{\rm charm}}(\theta_{{\cal E}E}= \pi)=\sqrt{ (g \omega_{i})^{2} \mathcal{E}^{2} 
                 + (eQ_{\rm charm})^{2} E^{2} 
                 - 2g \omega_{i}eQ_{\rm charm} \mathcal{E} E 
                 }$,  implying that pair production is most enhanced for the parallel configuration. This is true for any index of $i$ giving a positive eigenvalue $\omega_{i} > 0$. However, this eigenvalue appears with a partner $\omega_{j}$ having an opposite sign $\omega_j=-\omega_i$ [for SU(3) we have $\omega_1=-\omega_2=1/2$], and the antiparallel configuration gives the strongest effective field for the index $j$, $\bbb_{j,{\rm charm}}(\theta_{{\cal E}E}=\pi)=\bbb_{i,{\rm charm}}(\theta_{{\cal E}E}=0)$.  Therefore, after summing over all the pairwise modes $i$, we obtain the angle dependence symmetric with respect to $\theta_{{\cal E}E}=\pi/2$.

\begin{figure*}[t]
\begin{tabular}{cc}
\begin{minipage}{0.55\hsize}
\includegraphics[width=0.8 \textwidth, bb = 160 50 750 600]{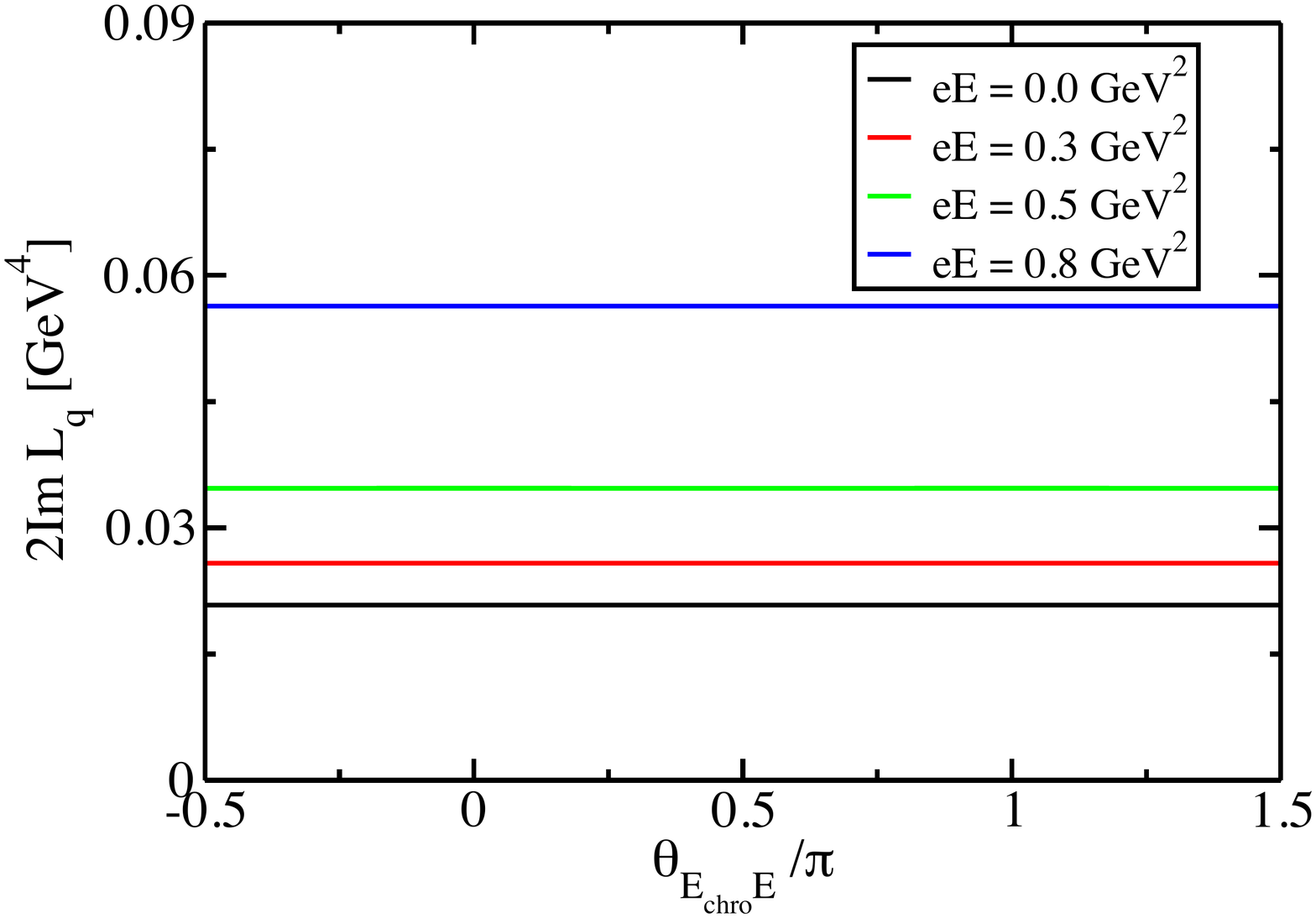}
\end{minipage}
\begin{minipage}{0.55\hsize}
\includegraphics[width=0.8 \textwidth, bb = 160 50 750 600]{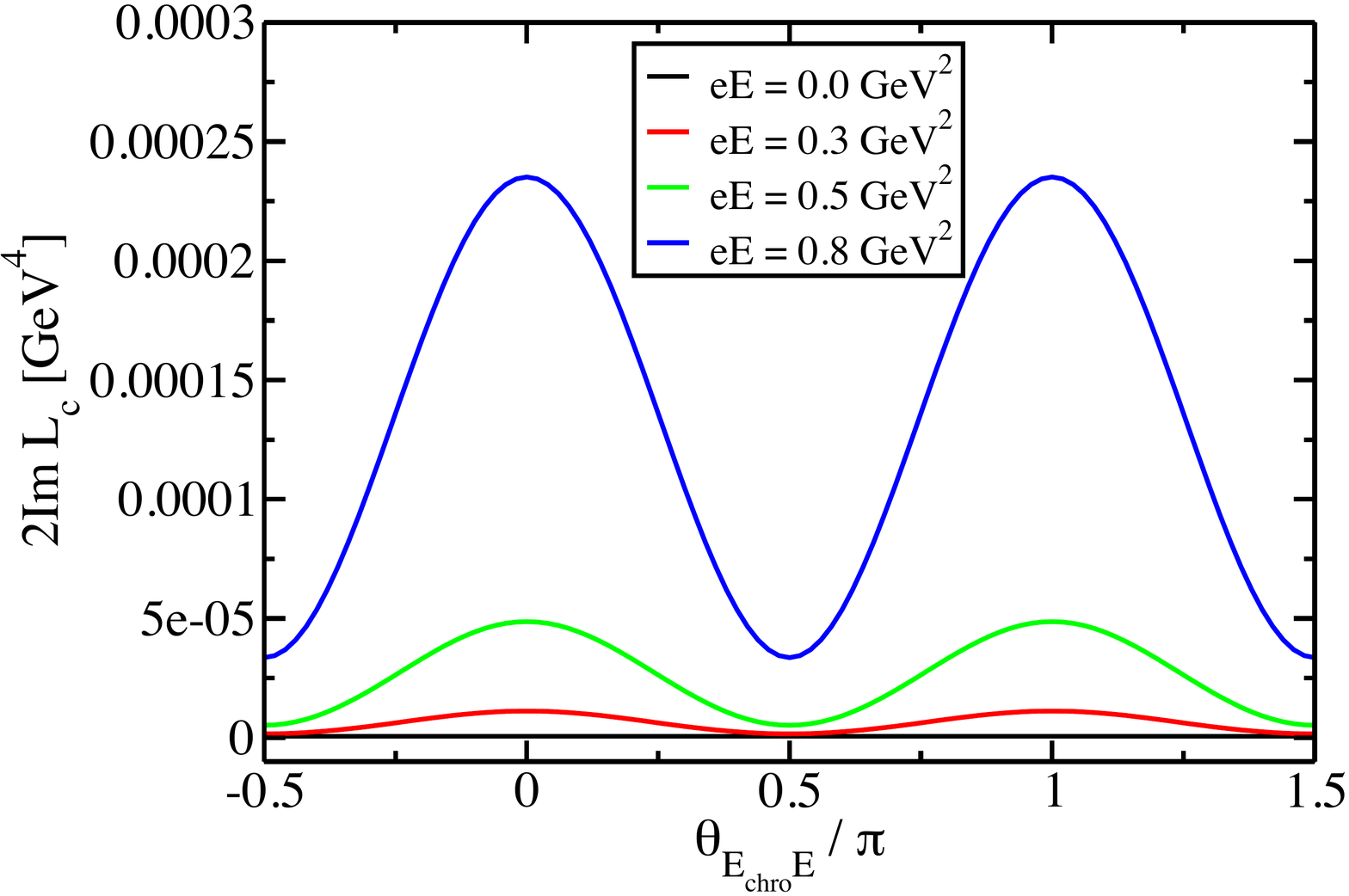}
\end{minipage}
\end{tabular}
\caption{ Quark production rate as a function of the angle $\theta_{E_{\rm{chro}}E}$, which stands for $\theta_{\mathcal{E}E}$.
The left panel is the light (up) quark production rate, while the right panel is the heavy (charm) quark production rate.
The chromo-electric field is fixed as $g\mathcal{E} = 1$ GeV$^{2}$. 
}
\end{figure*}

\subsubsection{Quark pair production in purely chromo-EM background}

Next, we investigate quark pair production under chromo-EM fields in the absence of EM fields. 
Lorentz-invariant quantities  $\mathbb{F}_{i,f}^{2}$ and $\mathbb{F}_{i,f}\cdot \tilde{\mathbb{F}}_{i,f}$ are now explicitly given as [see Eq.~(\ref{FFtilde})]
\beq
\mathbb{F}_{i,f}^{2} = 2 (g \omega_{i})^{2} (\mathcal{B}^{2} - \mathcal{E}^{2} )\, , \ \ \ \ \ \ 
\mathbb{F}_{i,f}\cdot \tilde{\mathbb{F}}_{i,f} = -4 (g \omega_{i})^{2} \mathcal{E} \mathcal{B} {\rm{cos}}\, \theta_{\mathcal{E} \mathcal{B} }\, ,
\eeq
where $\mathcal{B} = \sqrt{ \vec{ \mathcal{B} }^{2} }$, and $\theta_{\mathcal{E} \mathcal{B}}$ stands for the angle between $\vec{\mathcal{E}}$ and $\vec{\mathcal{B}}$. When $\theta_{\mathcal{E} \mathcal{B} } = \pm \pi/2$ and $\mathcal{E} > \mathcal{B}$, we can move into a system with pure chromo-electric fields with
$
\aaa_{i,f} = \aaa_{i} = 0 $ and
$\bbb_{i,f} = \bbb_{i} = |g \omega_{i}| \sqrt{\mathcal{E}^{2} - \mathcal{B}^{2} }$ by the Lorentz transformation. Then, the production rate for a certain flavor of quark becomes
\beq
2 {\Im m}\, \mathcal{L}_{\rm quark}
&=& \frac{1}{4\pi^{3}} \sum_{i=1}^{N_{c}} \bbb_{i}^{2} \sum_{n=1}^{\infty} \frac{1}{n^{2}} \, \e^{- \frac{ m_{q}^{2} }{\bbb_{i}} n \pi },
\eeq
which decreases as $\mathcal{B}$ increases. 
Furthermore, for $\mathcal{B} \ge \mathcal{E}$ the production rate vanishes since in this case the system is equivalent to the pure chromo-magnetic field system.
When $\theta_{\mathcal{E} \mathcal{B}} = 0, \pi$, which would be relevant configurations for relativistic heavy-ion collisions, $\aaa_{i}$ and $\bbb_{i}$ become
$
\aaa_{i} = | g \omega_{i} \mathcal{B}|$, $\bbb_{i} = |g \omega_{i} \mathcal{E}|$.
Then, the production rate reads
\beq
2 {\Im m}\, \mathcal{L}_{\rm quark}
&=& \frac{1}{4\pi^{2}} \sum_{i=1}^{N_{c}} | g \omega_{i} \mathcal{B} | | g \omega_{i} \mathcal{E}| \sum_{n=1}^{\infty} \frac{1}{n}\, \e^{ - \frac{ m_{q}^{2} }{ | g \omega_{i} \mathcal{E} | } n \pi }
{\rm{coth}} \left( \frac{ \mathcal{B} }{ \mathcal{E} } n \pi \right).
\label{chromoEB}
\eeq
This production rate is the same result as obtained in Refs.~\cite{Suganuma:1991ha, Tanji:2008ku}.
It increases as either the chromo-electric field or the chromo-magnetic field increases.
Figure~4 shows $\theta_{\mathcal{E} \mathcal{B}}$ dependence of the light quark production rate with a fixed value of the chromo-electric field, $g\mathcal{E}=1$ GeV$^{2}$.
The maxima appear when the chromo-magnetic field is parallel (or antiparallel) to the chromo-electric field.

\begin{figure}
\begin{minipage}{0.8\hsize}
\begin{center}
\includegraphics[width=0.8 \textwidth]{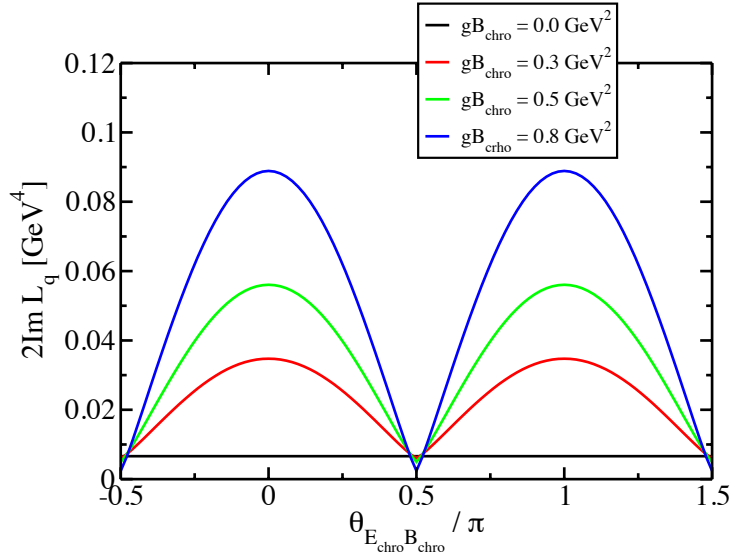}
\vskip -0.1in
\end{center}
\end{minipage}
\caption{
Light (up) quark production rate as a function of $\theta_{E_{\rm{chro}}B_{\rm{chro}}}$, which stands for $\theta_{\mathcal{E}\mathcal{B}}$ with vanishing electromagnetic fields.
We take the strength of the chromo-electric field as $g\mathcal{E} = 1$ GeV$^{2}$. 
}
\end{figure}

\subsubsection{Quark pair production in a glasma with EM fields}

Now we shall consider a specific configuration of chromo-EM fields that are relevant for relativistic heavy-ion collisions accompanied by EM fields. Suppose that the chromo-electric field and the chromo-magnetic field are parallel to each other, $\vec{\mathcal{B}} \parallel \vec{\mathcal{E}}$, and that these strengths are approximately equal to the saturation scale: $|g\vec{\mathcal{B}}| = |g\vec{\mathcal{E}}| = 1$~GeV$^{2} \sim Q_{s}^2$. 
This configuration of chromo-EM fields is indeed realized at the very early stage of the glasma evolution.
Under this condition, we investigate light (up) quark productions with $m_{q}=0.5$ MeV and $Q_{q} = +2/3$.
Let us turn on the EM fields. 
In the heavy-ion collisions, the dominant EM field is the magnetic field perpendicular to the beam direction (equivalent to the direction of the glasma fields). But here we consider the case $|e\vec{B}| \neq 0$ and $|e\vec{E}|=0$, with arbitrary orientation. Then, the quantities $\mathbb{F}_{i,f}^{2}$, and $\mathbb{F}_{i,f} \cdot \tilde{\mathbb{F}}_{i,f}$ read [see Eq.~(\ref{FFtilde})]
\beq
\mathbb{F}_{i,f}^{2}
&=& 2 \left[ (eQ_{q})^{2} B^{2} + 2 g \omega_{i} eQ_{q} \mathcal{B}B {\rm{cos}} \theta_{\mathcal{B}B} \right], \nonumber \\
\mathbb{F}_{i,f} \cdot \tilde{\mathbb{F}}_{i,f}
&=& -4 \left[ (g \omega_{i})^{2} \mathcal{E} \mathcal{B} + g \omega_{i} eQ_{q} \mathcal{E} B {\rm{cos}} \theta_{\mathcal{B}B} \right],
\eeq 
with $B = \sqrt{ \vec{B}^{2} }$. Here we have used the fact that ${\rm{cos}} \theta_{\mathcal{E} B} =  {\rm{cos}} \theta_{\mathcal{B}B}$. 
Note that in the case of antiparallel configuration of $\vec{\mathcal{B}}$ and $\vec{\mathcal{E}}$, results are the same as those of the parallel case, since this changes $\mathbb{F}_{i,f}\cdot \tilde{\mathbb{F}}_{i,f} \to - \mathbb{F}_{i,f}\cdot \tilde{\mathbb{F}}_{i,f}$, but it is squared in $\aaa_{i,f}$ and $\bbb_{i,f}$. 

Figure~5 shows the quark production rate as a function of the angle $\theta_{\mathcal{B}B}$ with several strengths of the magnetic field.
At the angle relevant for relativistic heavy-ion collisions, $\theta_{\mathcal{B} B} = \pi/2,$ the production rate slightly decreases with increasing $B$ field. This can be understood from Eq.~(\ref{ImLq_full}) as follows:
In this case, the quantity $\aaa_{i,f} = \frac{1}{2} \sqrt{ \sqrt{ 4(eQ_{q})^{4} B^{4} + 16 (g \omega_{i})^{4} \mathcal{E} \mathcal{B} } + 2(eQ_{q})^{2} B^{2} }$ (or $\bbb_{i,f} = \frac{1}{2} \sqrt{ \sqrt{ 4(eQ_{q})^{4} B^{4} + 16 (g \omega_{i})^{4} \mathcal{E} \mathcal{B} } - 2(eQ_{q})^{2} B^{2} }$ ) increases (decreases) with increasing $B$ field, while the product $\aaa_{i,f} \bbb_{i,f} = | \vec{ \mathcal{E} }_{i,f} \cdot \vec{ \mathcal{B} }_{i,f} |= (g \omega_{i})^{2} \mathcal{E} \mathcal{B}$ is independent of $B$ field.
Therefore, at $\theta_{\mathcal{B} B} = \pi/2$, the quark production rate monotonically decreases due to the exponential factor ${\rm exp}\{- (m_{q}^{2}/\bbb_{i,f}) n \pi\} $. 
This result is independent of the sign of $\omega_{i}$.

On the other hand, Fig.~5 shows that the quark production rate increases with increasing $B$ field at $\theta_{\mathcal{B}B} = 0$ and $\pi$. This can be understood as follows:
At $\theta_{\mathcal{B}B} = 0, \pi$, the quark production rate reads from Eq.~(\ref{ImLq_full})
\beq
2 {\Im m}\, \mathcal{L}_{\rm{quark}}
&=& \frac{1}{4\pi^{2}} \sum_{i=1}^{N_{c}} |g\omega_{i}| \mathcal{E} \mathcal{B}_{i,f} \sum_{n=1}^{\infty}
\frac{1}{n} \e^{- \frac{m_{q}^{2}}{|g\omega_{i}|\mathcal{E}} n \pi } \coth \left( \frac{ \mathcal{B}_{i,f} }{ |g\omega_{i}| \mathcal{E} } n \pi \right),
\label{choromoEBandB}
\eeq
where the strength of the combined magnetic field has been defined as $\mathcal{B}_{i,f} = |g\omega_{i} \mathcal{B} + eQ_{q}B | $ for $\theta_{\mathcal{B}B} = 0$, whereas $\mathcal{B}_{i,f} = |g\omega_{i} \mathcal{B} - eQ_{q}B | $ for $\theta_{\mathcal{B}B} = \pi$. 
This production rate has a similar form with Eq.~(\ref{chromoEB}).
First, we consider the case $|g\omega_{i} \mathcal{B}| > |eQ_{q}B|$. When the chromo-magnetic field and the magnetic field are (anti)parallel to each other, $\theta_{\mathcal{B}B} = 0$ ($\theta_{\mathcal{B}B} = \pi$), with $\omega_{i} > 0$ ($\omega_{i} < 0$), the strength of the combined magnetic field $\mathcal{B}_{i,f}$ linearly increases with increasing $B$ field, and thus $\coth \left( \frac{ \mathcal{B}_{i,f} }{ |g\omega_{i}| \mathcal{E} } n \pi \right)$ slightly decreases and approaches unity.
When $\theta_{\mathcal{B}B} = 0$ ($\theta_{\mathcal{B}B} = \pi$) with $\omega_{i} < 0$ ($\omega_{i} > 0$), the field strength $\mathcal{B}_{i,f}$ linearly decreases with increasing $B$ field, but $\coth \left( \frac{ \mathcal{B}_{i,f} }{ |g\omega_{i}| \mathcal{E} } n \pi \right)$ increases. Then, after summing over all the modes $i$, the production rate (\ref{choromoEBandB}) at $\theta_{\mathcal{B}B} = 0$ ($\theta_{\mathcal{B}B} = \pi$) monotonically increases with increasing $B$ field.
In the case of $|g\omega_{i} \mathcal{B}| \le |eQ_{q}B|$, the production rate of both modes $i=1,2$ increases with increasing $B$ field regardless of the sign of $\omega_{i}$, and thus the total production rate also monotonically increases.
Furthermore, we again obtain the angle dependence symmetric with respect to $\theta_{\mathcal{B}B} = \pi/2$ in the production rate.

\begin{figure}
\begin{minipage}{0.8\hsize}
\begin{center}
\includegraphics[width=0.8 \textwidth]{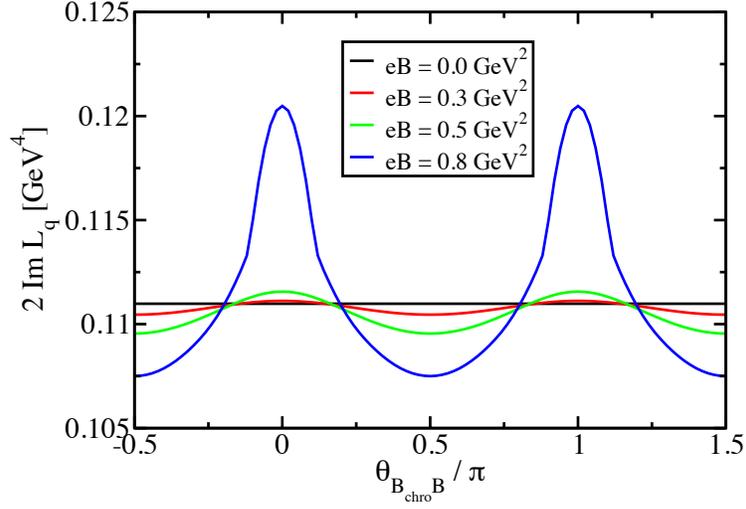}
\vskip -0.1in
\end{center}
\end{minipage}
\caption{
Light (up) quark production rate in a $B$ field as a function of $\theta_{B_{\rm{chro}B}}$, which stands for $\theta_{\mathcal{B} B}$ with a parallel configuration of $\vec{\mathcal{E}}$ and $\vec{\mathcal{B}}$. 
We take strengths of chromo-electromagnetic fields as $g\mathcal{B} = g \mathcal{E} = 1$~GeV$^{2}$.
}
\end{figure}

Next we consider the case with $|e\vec{E}| \neq 0$ and $|e\vec{B}|=0$.
In this case, $\mathbb{F}_{i,f}^{2}$ and $\mathbb{F}_{i,f} \cdot \tilde{\mathbb{F}}_{i,f}$ become [see Eq.~(\ref{FFtilde})]
\beq
\mathbb{F}_{i,f}^{2}
&=& 2 \left[- (eQ_{q_{f}})^{2} E^{2} - 2g \omega_{i}eQ_{q_{i}} \mathcal{E} E {\rm{cos}} \theta_{\mathcal{E} E} \right], \nonumber \\
\mathbb{F}_{i,f} \cdot \tilde{\mathbb{F}}_{i,f}
&=& -4 \left[ (g \omega_{i})^{2} \mathcal{E} \mathcal{B} + g \omega_{i}eQ_{q_{f}} \mathcal{B} E {\rm{cos}} \theta_{\mathcal{E}E} \right]\, .
\eeq
In this expression, we have used ${\rm{cos}} \theta_{\mathcal{B}E} = {\rm{cos}} \theta_{\mathcal{E}E}$.
Again, the results are the same as those of the case where $\vec{\mathcal{B}}$ is antiparallel to $\vec{\mathcal{E}}$.
Figure~6 shows the quark production rate as a function of the angle $\theta_{\mathcal{E}E}$ with several values of strength of the electric field.
As the electric field increases, the production rate increases for whole angle regions. 
This can be understood in a similar way to the previous case as follows:
At $\theta_{\mathcal{E} E} = \pi/2$, the factor $\aaa_{i,f} \bbb_{i,f} = | \vec{ \mathcal{E} }_{i,f} \cdot \vec{ \mathcal{B} }_{i,f} |= (g \omega_{i})^{2} \mathcal{E} \mathcal{B}$ is independent of the electric field.
As for each factor, $\aaa_{i,f} = \frac{1}{2} \sqrt{ \sqrt{ 4(eQ_{q})^{4} E^{4} + 16 (g \omega_{i})^{4} \mathcal{E} \mathcal{B} } - 2(eQ_{q})^{2} E^{2} }$ decreases with increasing electric field, while $\bbb_{i,f} = \frac{1}{2} \sqrt{ \sqrt{ 4(eQ_{q})^{4} E^{4} + 16 (g \omega_{i})^{4} \mathcal{E} \mathcal{B} } + 2(eQ_{q})^{2} E^{2} }$ increases. 
These behaviors are opposite to those of the previous case with $|e\vec{E}| = 0$ and $|e\vec{B}| \neq 0$, and thus the production rate at $\theta = \pi/2$ monotonically increases. At $\theta_{\mathcal{E} E} = 0, \pi$,  the quark production rate (\ref{ImLq_full}) can be rewritten as
\beq
2 {\Im m}\, \mathcal{L}_{\rm{quark}}
&=& \frac{1}{4\pi^{2}} \sum_{i=1}^{N_{c}} \mathcal{E}_{i,f} |g\omega_{i}|  \mathcal{B} \sum_{n=1}^{\infty}
\frac{1}{n} \e^{- \frac{m_{q}^{2}}{\mathcal{E}_{i,f}} n \pi } \coth \left( \frac{ |g\omega_{i} |\mathcal{B} }{ \mathcal{E}_{i,f} } n \pi \right),
\label{chromoEBandE}
\eeq
where the strength of the combined electric field has been defined as $\mathcal{E}_{i,f} = | g\omega_{i} \mathcal{E} + e Q_{q}E| $ for $\theta_{\mathcal{E} E} = 0$ and $\mathcal{E}_{i,f} = | g\omega_{i} \mathcal{E} - e Q_{q}E| $ for $\theta_{\mathcal{E} E} = \pi$. 
In the case of $|g\omega_{i} \mathcal{E}| > |e Q_{q}E|$, when the chromo-electric field and the electric field are (anti)parallel to each other, $\theta_{\mathcal{E} E} = 0$ ($\theta_{\mathcal{E} E} = \pi$), with $\omega_{i} > 0$ ($\omega_{i} < 0$), the strength of the combined electric field $\mathcal{E}_{i,f}$ linearly increases with increasing $E$ field, and thus $\coth \left( \frac{ |g\omega_{i} |\mathcal{B} }{ \mathcal{E}_{i,f} } n \pi \right)$ monotonically increases.
When $\theta_{\mathcal{E} E} = 0$ ($\theta_{\mathcal{E} E} = \pi$) with $\omega_{i} < 0$ ($\omega_{i} > 0$), the field strength $\mathcal{E}_{i,f}$ linearly decreases with increasing $E$ field, and $\coth \left( \frac{ |g\omega_{i} |\mathcal{B} }{ \mathcal{E}_{i,f} } n \pi \right)$ slightly decreases and approaches unity. 
Then, after summing over all the modes $i$, the production rate (\ref{chromoEBandE}) at $\theta_{\mathcal{E} E} = 0$ ($\theta_{\mathcal{E} E} = \pi$) monotonically increases with increasing $E$ field.
On the other hand, in the case of $|g\omega_{i} \mathcal{E}| \le |e Q_{q}E|$, the production rate of both modes $i=1,2$ increases with increasing $E$ field regardless of the sign of $\omega_{i}$, and thus the total production rate also monotonically increases.
From these results, we expect that strong EM fields created in the early stage of relativistic heavy-ion collisions would largely affect quark productions from a glasma (chromo-EM fields) depending on the field configurations, and would thus possibly influence the formation of QGP.

\begin{figure}
\begin{minipage}{0.8\hsize}
\begin{center}
\includegraphics[width=0.8 \textwidth]{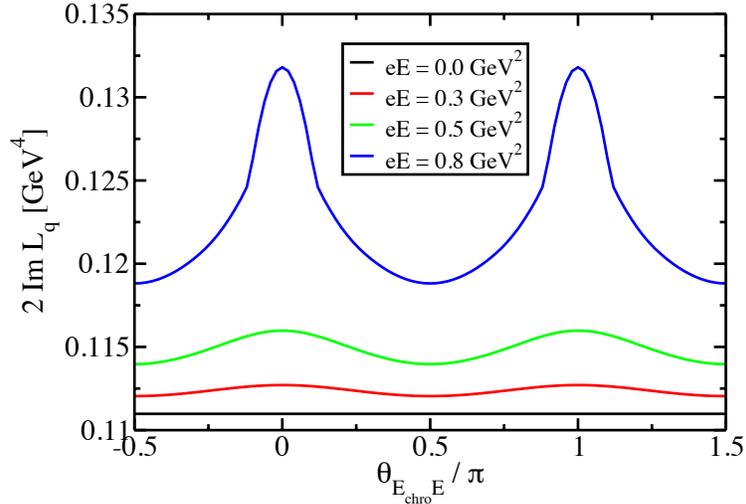}
\vskip -0.1in
\end{center}
\end{minipage}
\caption{
Light (up) quark production rate in an $E$ field as a function of $\theta_{E_{\rm{chro}}E}$, which stands for $\theta_{\mathcal{E} E}$ with a parallel configuration of $\vec{\mathcal{E}}$ and $\vec{\mathcal{B}}$. 
We take strengths of chromo-electromagnetic fields as $g\mathcal{B} = g \mathcal{E} = 1$ GeV$^{2}$.
}
\end{figure}

\subsection{Weiss potential with electromagnetic fields}

In this subsection, we will investigate the effects of EM fields on the confinement-deconfinement phase transition by using the effective potential of the Polyakov loop in the presence of EM fields. 

Prior to going into the details, let us briefly explain the effective potential without external fields being imposed. The one-loop calculation at finite temperature in SU(2) gauge theory and in the massless fermion limit yields the effective potential for the temporal component of the gauge field $(C = \frac{ g \bar{\A}_{4} }{ 2 \pi T })$ as~\cite{Weiss:1980rj,Weiss:1981ev, Gross:1980br}
\beq
V^{\rm Weiss}[C]=V_{\rm YM}^{\rm Weiss}[C]+V^{\rm Weiss}_{\rm quark}[C]\, ,
\eeq
where the YM and quark parts are given, respectively, by  
\beq
V_{\rm YM}^{\rm Weiss}[C]&=&- \frac{ 3 }{ 45 } \pi^{2} T^{4} + \frac{3}{4} \pi^{2} T^{4} C^{2} (1-C)^{2}\, ,\label{Weiss_YM}\\
V_{\rm quark}^{\rm Weiss}[C]&=&- \frac{7}{90} \pi^{2} T^{4} + \frac{1}{6} \pi^{2} T^{4} C^{2} ( 2 - C^{2} )\, .\label{Weiss_quark}
\eeq
This result is called the Weiss potential. In Fig.~7, we show the Weiss potential $V^{\rm Weiss}[C]$ and its breakdown. 
We see that in the YM part, the minima appear at $C=0$ and $C = 1$, reflecting the center symmetry $C\to C+1$ in SU(2). Thus, selecting one of the two minima spontaneously breaks the center symmetry.
Since the system should be in the deconfined phase in the high-temperature region where a perturbative approach becomes valid,
this result seems to be natural. 
The quark part of the effective potential explicitly breaks the center symmetry, and $C=0$ and $C=1$ are no longer degenerated. 
In the presence of the quark part, $C=0$ is favored, which corresponds to the deconfined phase. We are now going to investigate how this picture is modified by the presence of external EM fields.
\begin{figure}
\begin{minipage}{0.8\hsize}
\begin{center}
\includegraphics[width=0.8 \textwidth]{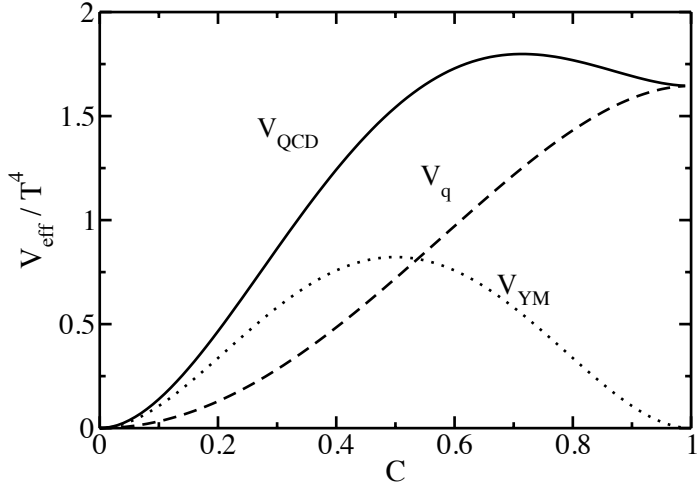}
\vskip -0.1in
\end{center}
\end{minipage}
\caption{Weiss potential as a function of $C$. 
Constant terms which are independent of $C$ are subtracted.
}
\end{figure}

Now we come back to our most general results (\ref{action_gluon}), (\ref{action_ghost}), and (\ref{action_quark}). 
Taking the vanishing limit of the chromo-EM fields, $\vec{\mathcal{E}}, \vec{\mathcal{B}} \to 0$, but keeping the Polyakov loop $\bar{\A}_{4}$ and EM fields nonzero in the results , we obtain the effective potential
\beq
V_{\rm eff} [\bar{\A}_{4}, E, B]
&=& - \frac{S_{\rm eff}}{\int dx^{4} } \nonumber \\
&=& \frac{1}{32 \pi^{2}} \sum_{h=1}^{N_{c}^{2}-1} \int^{\infty}_{0} \frac{ds}{s^{3}} \left\{ 4-2 \right\} 2 \sum_{n=1}^{\infty} \e^{ i\frac{n^{2}}{4T^{2}s} }
{\rm{cos}} \left(  \frac{ g v_{h} \bar{\A}_{4} }{T} n \right) \nonumber \\
&& - \frac{1}{8\pi^{2}} \sum_{i=1}^{N_{c}} \sum_{f=1}^{N_{f}} \int^{\infty}_{0} \frac{ds}{s^{3}} \e^{-im_{q_{f}}^{2}s} (\aaa_{f}s)( \bbb_{f}s ){\rm{cot}}(\aaa_{f}s) {\rm{coth}}(\bbb_{f}s) \nonumber \\
&& \times 2 \sum_{n=1}^{\infty} (-1)^{n} \e^{i \frac{1}{4T^{2}} \mathfrak{h}_{f}(s) n^{2}} {\rm{cos}} \left( \frac{ g \omega_{i} \bar{\A}_{4} }{T} n \right).
\eeq
where $\aaa_f$ and $\bbb_f$ are just given by the EM fields as
\beq
\aaa_{f}
= \frac{1}{2} \sqrt{ \sqrt{ F_{f}^{4} + (F_{f}\cdot \tilde{F}_{f})^{2} } + F_{f}^{2} }\, , \qquad 
\bbb_{f}
= \frac{1}{2} \sqrt{ \sqrt{ F_{f}^{4} + (F_{f} \cdot \tilde{F}_{f} )^{2} } - F_{f}^{2} }\, ,
\eeq
with $F_{f}^{2} = 2 (eQ_{q_{f}})^{2} ( \vec{B}^{2} - \vec{E}^{2} ) $ and $F_{f} \cdot \tilde{F}_{f} = -4 (eQ_{q_{f}})^{2} \vec{E} \cdot \vec{B}$. 
The factor $\mathfrak{h}_{f}(s)$ is given by
\beq
\mathfrak{h}_{f}(s)
&=& \frac{ \bbb_{f}^{2} - {\eee}_{f}^{2} }{ \aaa_{f}^{2} + \bbb_{f}^{2} } \aaa_{f} {\rm{cot}}( \aaa_{f}s ) + \frac{ \aaa_{f}^{2} + {\eee}_{f}^{2} }{ \aaa_{f}^{2} + \bbb_{f}^{2} } \bbb_{f} {\rm{coth}}(\bbb_{f} s)\, ,
\eeq
where ${\eee}_{f}^{2} = (u_{\alpha} F_{f}^{\alpha \mu})( u_{\beta} F_{f \mu}^{\beta}) = (eQ_{q_{f}})^{2} E^{2} $ with $u_{\mu} = (1,0,0,0)$.
Here we have subtracted divergences appearing in the zero-temperature contribution, which are independent of 
$\bar{\A}_{4}$.

\subsubsection{Weiss potential in magnetic fields}

Consider a pure magnetic field case, $\vec{E} \to 0$, $\vec{B} \neq 0$.
Then, the effective potential reads,
\beq
V_{\rm eff} [\bar{\A}_{4}, B]
&=&\frac{1}{32 \pi^{2}} \sum_{h=1}^{N_{c}^{2}-1} \int^{\infty}_{0} \frac{ds}{s^{3}} \left\{ 4-2 \right\} 2 \sum_{n=1}^{\infty} \e^{ i\frac{n^{2}}{4T^{2}s} }
\, {\rm{cos}} \left(  \frac{ g v_{h} \bar{\A}_{4} }{T} n \right) \nonumber \\
&& - \frac{1}{8\pi^{2}} \sum_{i=1}^{N_{c}} \sum_{f=1}^{N_{f}} \int^{\infty}_{0} \frac{ds}{s^{3}} \e^{-im_{q_f}^{2}s} ( e|Q_{q_{f}}|B s) {\rm{cot}}( e|Q_{q_{f}}|Bs) \nonumber \\
&&\qquad \times 2 \sum_{n=1}^{\infty} (-1)^{n} \e^{i \frac{ n^{2} }{ 4T^{2}s } }\, {\rm{cos}} \left( \frac{ g \omega_{i}\bar{\A}_{4} }{T} n \right).
\label{VB}
\eeq 
We rewrite the proper time integrals in two steps. Recall that the integral should be defined with an infinitesimally small number $\delta$ which makes the contour slightly inclined to avoid the poles along the real axis (in the second term). Then we can easily change the contour from $[0,\infty]$ along the real axis to $[-i\infty,0]$ along the imaginary axis (the Wick rotation), since there is no pole along the imaginary axis. Finally, by renaming the variable $s$ as $-i\sigma$, we obtain the following representation with integrals defined by real functions\footnote{The second line of Eq.~(\ref{PolyakovLoopwithB}) coincides with Eq.~(B.6) in the appendix of Ref.~\cite{Bruckmann:2013oba}.}:
\beq
V_{\rm eff} [\bar{\A}_{4}, B]
&=& - \frac{ 1 }{ 8\pi^{2}} \sum_{h=1}^{N_{c}^{2}-1} \int^{\infty}_{0} \frac{d\sigma }{\sigma^{3}} \sum_{n=1}^{\infty} \e^{- \frac{n^{2}}{4T^{2}\sigma} }\, {\rm{cos}}\left( \frac{ g v_{h} \bar{\A}_{4} }{ T } n \right) \nonumber \\
&& +\frac{1}{4\pi^{2}} \sum_{i=1}^{N_{c}} \sum_{f=1}^{N_{f}} \int^{\infty}_{0} \frac{d\sigma}{\sigma^{2}} \e^{-m_{q_{f}}^{2}\sigma} (e|Q_{q_{f}}|B) {\rm{coth}}(e|Q_{q_{f}}|B\sigma ) \nonumber \\
&&\qquad \times \sum_{n=1}^{\infty} (-1)^{n} \e^{ - \frac{n^{2}}{4T^{2}\sigma} }\, {\rm{cos}} \left( \frac{ g \omega_{i}\bar{\A}_{4} }{ T} n \right).
\label{PolyakovLoopwithB}
\eeq
For simplicity, we shall restrict ourselves to $N_{c}=2$, which provides us with all the essential features of the perturbative effective potential in the presence of EM fields. In this case, the eigenvalues $\omega_{i}$ and $v_{h}$ are simply given by $\omega_{i} = \pm 1/2$ and $v_{h} = 0, \pm 1$.
The effective potential reads,
\beq
V_{\rm eff}[ C, B ]
&=& - \frac{ 3 }{ 45 } \pi^{2} T^{4} + \frac{3}{4} \pi^{2} T^{4} C^{2} (1-C)^{2}  \label{effectiveVTB} \\
&& + \frac{1}{2\pi^{2}} \sum_{f=1}^{N_{f}} \int^{\infty}_{0} \frac{d\sigma }{\sigma^{2}} \e^{-m_{q_{f}}^{2}\sigma} (e|Q_{q_{f}}|B) {\rm{coth}}(e|Q_{q_{f}}|B\sigma ) \sum_{n=1}^{\infty} (-1)^{n} \e^{ - \frac{n^{2}}{4T^{2}\sigma} } {\rm{cos}} \left( C\pi n \right)\, . \nonumber 
\eeq
The first line does not depend on the magnetic field and corresponds to the YM part $V_{\rm YM}$. This is nothing but the Weiss potential~(\ref{Weiss_YM}) \cite{Weiss:1980rj}. The second line corresponds to the quark part $V_{\rm quark}$, and the integral and summation over $n$ can be easily performed numerically. 
From now on, we further restrict ourselves to the one flavor $f=1$ with the electric charge $Q_{q_{f}}=1$ for simplicity.
Now, analytic expressions are available in two limiting cases: One is the $B\to 0$ and $m_q\to 0$ limit, where the quark part of the effective potential is reduced to that of the Weiss potential (\ref{Weiss_YM}):
\beq
V_{\rm quark} [C]
&= & - \frac{7}{90} \pi^{2} T^{4} + \frac{1}{6} \pi^{2} T^{4} C^{2} ( 2 - C^{2} )=V^{\rm Weiss}_{\rm quark}[C]\, .
\eeq
The other is the strong magnetic field limit: $eB \gg m_{q}^{2}$, where the quark part can be written as
\beq
V_{\rm quark} [C, B]
&=& - 2 \frac{ (eB) }{ \pi^{2} } T^{2} \left\{ \frac{ \pi^{2} }{ 12 } - \frac{ (C\pi)^{2} }{ 4 } \right\}. \label{quarkVstrongB}
\eeq
Figure~8 shows the magnetic field dependence of the quark part of the effective potential which is given by the second line of Eq.~(\ref{effectiveVTB}). 
Here, we show only one flavor contribution with $x=m_q^2/T^2=0.5$. An important observation is that as the magnetic field increases, the explicit breaking of the center symmetry is enhanced, and $C=0$ (deconfined phase) becomes more stable. This is qualitatively consistent with the analytic representation at strong magnetic fields [see Eq.~(\ref{quarkVstrongB})] in that the potential value at $C=0$ becomes more negative and the rising behavior becomes steeper with increasing magnetic field. The enhancement of the center symmetry-breaking effects due to increasing magnetic field indicates that the quark loop interacting with magnetic fields can be one of the important sources for reducing the (pseudo)critical temperature $T_{c}$ of confinement-deconfinement phase transition, as observed in recent lattice QCD simulations \cite{Bruckmann:2013oba}. In the last part of this subsection, we will see within a phenomenological model that this is indeed the case.

\begin{figure}[t]
\begin{minipage}{0.8\hsize}
\begin{center}
\includegraphics[width=0.8 \textwidth]{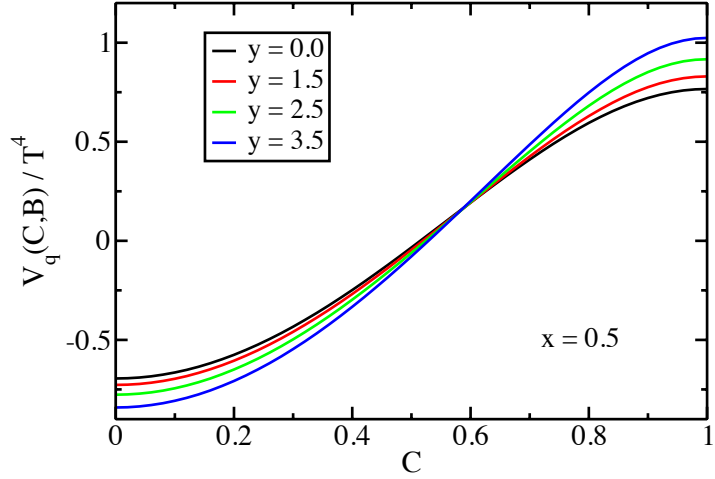}
\end{center}
\end{minipage}
\caption{Quark part of the effective potential as a function of $C$ for several values of magnetic fields. $x$ and $y$ are given as $x = m_{q}^{2}/T^{2}$ and $y = eB/T^{2}$, respectively. 
}
\end{figure}

\subsubsection{Weiss potential in electric fields}

In the case of a pure electric field, $\vec{B} \to 0$ and $\vec{E} \neq 0$, the situation is a bit subtle. 
The effective potential of the quark part can be written as
\beq
V_{\rm quark}[\bar{\A}_{4}, E]
&=& - \frac{ 1}{2\pi^{2}} \sum_{f=1}^{N_{f}} \int^{\infty}_{0} \frac{ds}{s^{3}} \e^{-im_{q_{f}}^{2}s} \left( e|Q_{q_{f}}|Es \right) {\rm{coth}} \left( e| Q_{q_{f}}| Es \right) \nonumber \\
&& \quad \times \sum_{n=1}^{\infty} (-1)^{n} \e^{i \frac{n^2}{4T^{2}s} \left( e |Q_{q_{f}}| Es \right) {\rm{coth}} \left( e |Q_{q_{f}}|Es \right) } {\rm{cos}} \left( \frac{ g\bar{\A}_{4} }{2T } n \right).
\eeq
Note that we cannot reach this result from Eq.~(\ref{VB}) by replacing $B$ with $iE$, unlike the zero-temperature contribution. This is due to the form of the factor $\mathfrak{h}_{f}(s)=(e|Q_{q_f}|Es){\rm coth}(e|Q_{q_f}|Es)$ in the exponential. Because of this factor, the full calculation (even numerical evaluation) is rather difficult. Furthermore, since there are singularities (poles) on the imaginary axis, we cannot perform the Wick rotation of the proper time $s$, unlike the Weiss potential in magnetic fields. 
To avoid these difficulties, we expand the effective potential with respect to the electric field. Using $x{\rm{coth}}x \sim 1 + x^{2}/3 \cdots$, we get
\beq
V_{\rm quark}[\bar{\A}_{4}, E]
&=& - \frac{1}{2\pi^{2}} \sum_{f=1}^{N_{f}} \int^{\infty}_{0} \frac{ds}{s^{3}} \e^{-im_{q_{f}}^{2}s } \sum_{n=1}^{\infty} (-1)^{n} \e^{i \frac{ n^{2} }{4T^{2}s} } {\rm{cos}} \left( \frac{ g \bar{\A}_{4} }{2T } n \right) \nonumber \\
&& - \frac{ 1}{6 \pi^{2}} \sum_{f=1}^{N_{f}} (e|Q_{q_{f}}| E)^{2} \int^{\infty}_{0} \frac{ds}{s} \e^{-im_{q_{f}}^{2}s} \sum_{n=1}^{\infty} (-1)^{n} \e^{i \frac{ n^{2}}{4T^{2}s} } \left( 1 + \frac{ n^{2} }{ 4 T^{2} s } \right) {\rm{cos}} \left( \frac{ g \bar{\A}_{4} }{ 2T } n \right) \nonumber \\
&&  + {\cal O}(E^{4})\, .
\eeq
At this stage, we can perform the Wick rotation for the proper time $s$. Then, the effective potential reads
\beq
V_{\rm quark}[C, E]
&=& 
\frac{1}{2\pi^{2}} \sum_{f=1}^{N_{f}} \int^{\infty}_{0} \frac{d\sigma}{\sigma^{3}} \e^{-m_{q_{f}}^{2}\sigma} \sum_{n=1}^{\infty} (-1)^{n} \e^{- \frac{n^{2}}{4T^{2}\sigma} } {\rm{cos}}  \left( 
 C \pi n \right) \nonumber \\
 && - \frac{1}{6\pi^{2}} \sum_{f=1}^{N_{f}} (e|Q_{q_{f}}|E )^{2} \int^{\infty}_{0} \frac{d\sigma}{\sigma} \e^{-m_{q_{f}}^{2}\sigma} 
 \sum_{n=1}^{\infty} (-1)^{n} \e^{- \frac{ n^{2} }{ 4T^{2} \sigma } } \left( 1 - \frac{ n^{2} }{ 4 T^{2} \sigma } \right) {\rm{cos}} \left( C \pi n \right) \nonumber \\
 && + {\cal O}(E^{4})\, .
\eeq
The systematic expansion with respect to the $E$ field is possible, and the integral and sum can be performed numerically at each order.

In Fig. 9  we show the electric field dependence of the quark part of the effective potential. From this figure, we see that the electric field decreases the explicit breaking of the center symmetry. This is completely opposite to the $B$ dependence of the effective potential. Thus, we expect that $T_{c}$ increases with increasing $E$ field and approaches the $T_{c}$ of the pure YM theory.

\begin{figure}[t] 
\begin{minipage}{0.8\hsize}
\begin{center}
\includegraphics[width=0.8 \textwidth]{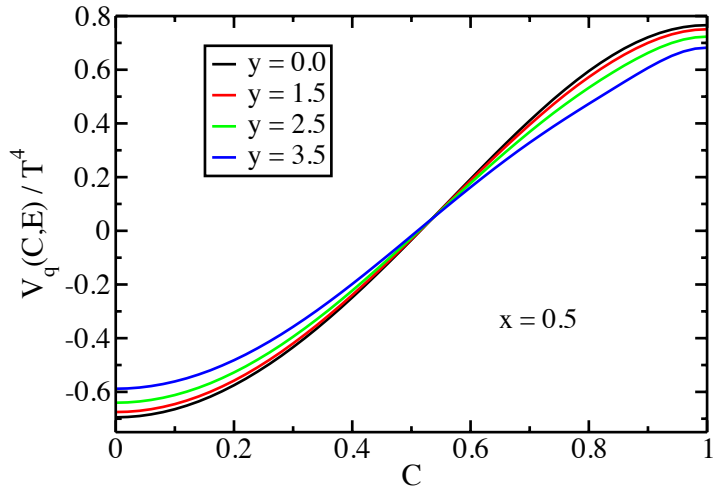}
\end{center}
\end{minipage}
\caption{ Quark part of the effective potential as a function of $C$ for several values of electric fields. $x$ and $y$ are given as $x = m_{q}^{2}/T^{2}$ and $y = eE/T^{2}$, respectively. 
}
\end{figure}

\subsubsection{Phenomenological analysis on $T_c(B)$}

We have seen that imposing magnetic fields enhances the explicit breaking of the center symmetry. What we have evaluated is a perturbative contribution (in the sense that we assume that the coupling is small enough), and thus we discussed how the Weiss potential (that is also evaluated in a perturbative framework) is modified in the presence of the EM fields. Within this perturbative calculation, we are not able to approach the region where phase transition will take place. Indeed, even if the quark part of the effective potential depends on the magnetic fields $V_{\rm quark}[C,B]$, the total effective potential $V_{\rm eff}[C,B]=V_{\rm YM}[C]+V_{\rm quark}[C,B]$ selects the center broken state $C=0$, and thus confinement-deconfinement phase transition never occurs within this perturbative framework. However, recall that the magnetic field can affect the effective potential of the Polyakov loop only through the quark loop at leading order. Therefore, we expect that even the perturbative evaluation of the quark part $V_{\rm quark}[C,B]$ can make sense if combined with some nonperturbative effective potential $V_{\rm YM}^{\rm nonpert}[C]$ for study of the effects of magnetic fields on the phase transition. Here we discuss whether this is indeed the case.

Let us introduce a simple model of a gluonic potential reproducing confinement-deconfinement phase transition,
\beq
\mathcal{U}[C]
&=& -\frac{1}{2}a(T) \Phi^{2} + b(T)\, {\rm{ln}} \left[ 1 - 6 \Phi^{2} + 8 \Phi^{3} - 3 \Phi^{4} \right]
\label{phenomenological_potential}
\eeq
with
\beq
a(T) = a_{0} + a_{1}(T_{0}/T) + a_{2}(T_{0} / T )^{2}, \ \ \ \ b(T) = b_{3}(T_{0}/T)^{3}.
\eeq
Now, we consider the $N_{c}=3$ case.
Here the parameters are
$a_{0} = 3.51,\, a_{1} = -2.47,\, a_{2} = 15.2,\, b_{3} = -1.75$, and $T_{0} = 270$ MeV, which are fixed to reproduce the quenched lattice QCD results \cite{Roessner:2006xn}.
Instead of $V_{YM}$, we employ this phenomenological potential (\ref{phenomenological_potential}) and combine it with $V_{\rm quark} [C,B]$. In this way, we can study how the temperature dependence of the Polyakov loop changes with magnetic fields. 
Notice that the quark part of the perturbative effective potential  $V_{\rm quark}[C,B]$ with $N_{c}=3$ is the same as that of the one with $N_{c}=2$, since the quark with $\omega_{3}=0$ does not contribute to the potential.
Therefore, we can use the same potential evaluated in the second line of Eq.~(\ref{effectiveVTB}).
The result is shown in Fig. 10. In this analysis, we have used $\omega_{i} = \pm 1/2, 0$ and a constituent quark mass $m_{q} = 350$ MeV. Thanks to the explicit center symmetry breaking, the Polyakov loop increases with increasing $B$ field, in particular below the phase transition temperature, 
which eventually brings about decreasing pseudocritical temperature $T_c(B)<T_c(B=0)$. This result is very encouraging, but obviously we need to couple quark dynamics to the gluon dynamics to understand the effects of magnetic fields on the actual phase transition.

Very recently, the inverse magnetic catalysis of the chiral sector, namely the decrease of the critical temperature of the chiral phase transition, has been reproduced from functional approaches including the Dyson-Schwinger equations and the functional renormalization group~\cite{Braun:2014fua, Mueller:2015fka}. Once the inverse magnetic catalysis of the chiral sector occurs, dynamical quark masses decrease with increasing magnetic field around $T_{c}$. 
Then, the quark loop contribution is enhanced, and thus the effect of the explicit center symmetry breaking becomes larger. Therefore, the inverse magnetic catalysis of chiral sector would support the decreasing of the $T_{c}$ of confinement-deconfinement phase transition through the quark loop.

\begin{figure}
\begin{minipage}{0.8\hsize}
\begin{center}
\includegraphics[width=0.8 \textwidth]{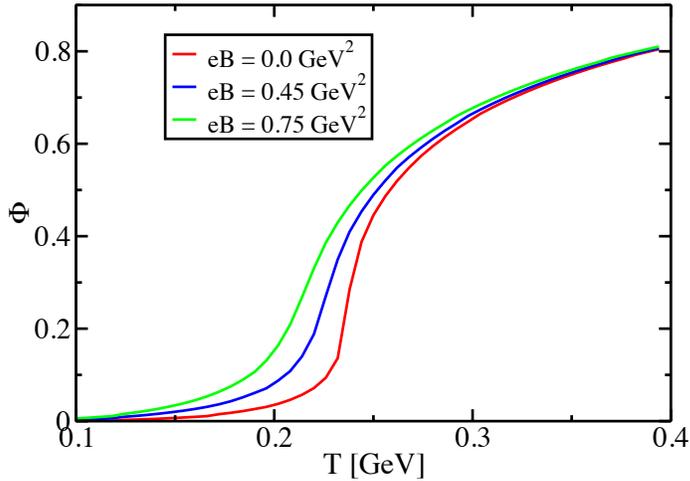}
\vskip -0.1in
\end{center}
\end{minipage}
\caption{
Temperature dependence of the Polyakov loop for different values of magnetic field.
}
\end{figure}

\section{Summary and conclusion}

In the present paper, we analytically derived the Euler-Heisenberg action for QCD+QED in the presence of the Polyakov loop, called the Euler-Heisenberg-Weiss action, by using the Schwinger proper time method. The effective action contains EM fields and chromo-EM fields as well as the Polyakov loop in a nonlinear form and reproduces the known one-loop effective actions for QED, QCD, QCD+QED, and also the Weiss potential for the Polyakov loop in appropriate limits. 

As an application of our effective action, we investigated quark pair productions under strong EM fields and chromo-EM fields. Using the effective action of the quark part at zero temperature, we derived the formula describing the quark pair production rate in arbitrary configurations of the QCD fields and QED fields.  In particular configurations, EM fields enhance the quark pair productions induced by chromo-EM fields. This indicates that strong EM fields created in relativistic heavy-ion collisions would largely affect quark-antiquark pair productions from a glasma and thus could give sizable contributions to the formation of QGP.

We also studied the perturbative effective action of the Polyakov loop in the presence of strong EM fields. We found that the magnetic (electric) field enhances (reduces) the explicit center symmetry breaking through the quark loop. This indicates that the Polyakov loop increases as the magnetic field increases, and thus the (pseudo)critical temperature of confinement-deconfinement phase transition decreases. In contrast, the electric field would raise the critical temperature. In order to demonstrate this, we combined  the quark part of our perturbative effective potential with a simple model which can reproduce the confinement-deconfinement phase transition. The resultant Polyakov loop indeed increases with increasing $B$ field, and then (pseudo)critical temperature decreases. This result is consistent with recent lattice data. 
Very recently, G.~Endrodi investigated QCD phase transitions in unprecedentedly strong magnetic fields from lattice simulations of 1 + 1 + 1-flavor QCD \cite{Endrodi:2015oba}.
He found strong evidence for a first-order confinement-deconfinement phase transition in the asymptotically strong magnetic field regions.
In order to understand these lattice data, further nonperturbative analyses will be necessary.
As a future work, we will extend the present work to nonperturbative analyses in terms of functional approaches.
The inclusion of the chiral sector (quark-quark interaction mediated by gluons) will also be an important ingredient in the future work.

\section*{Acknowledgements}

This work was supported in part by the Center for the
Promotion of Integrated Sciences (CPIS) of Sokendai.
The research of K.H. is supported by JSPS Grant-in-Aid No.~25287066.


\end{document}